\documentclass[journal]{IEEEtran}

\usepackage{cite}      

\usepackage{graphicx}  

\begin{document}

\title{Oxide spintronics}

\author{Manuel Bibes \thanks{M.B. is with the Institut d'Electronique Fondamentale, CNRS, Universit\'e Paris-Sud, 91405 Orsay, France. Email : manuel.bibes@ief.u-psud.fr}%
and Agn\`es Barth\'el\'emy \thanks{A.B. is with the Unit\'e Mixte de Physique CNRS-Thales, route
d\'epartementale 128, 91767 Palaiseau, France. Email: agnes.barthelemy@thalesgroup.com}}

\markboth{IEEE Trans. Electron. Devices,~Vol.~x, No.~xx,~August~2006}{Bibes and Barth\'el\'emy : Oxide
spintronics}
%




\maketitle

\begin{abstract}
Concomitant with the development of metal-based spintronics in the late 1980's and 1990's, important advances
were made on the growth of high-quality oxide thin films and heterostructures. While this was at first motivated
by the discovery of high-temperature superconductivity in perovskite Cu oxides, this technological breakthrough
was soon applied to other transition metal oxides, and notably mixed-valence manganites. The discovery of
colossal magnetoresistance in manganite films triggered an intense research activity on these materials, but the
first notable impact of magnetic oxides in the field of spintronics was the use of such manganites as electrodes
in magnetic tunnel junctions, yielding tunnel magnetoresistance ratios one order of magnitude larger than what
had been obtained with transition metal electrodes. Since then, the research on oxide spintronics has been
intense with the latest developments focused on diluted magnetic oxides and more recently on multiferroics. In
this paper, we will review the most important results on oxide spintronics, emphasizing materials physics as
well as spin-dependent transport phenomena, and finally give some perspectives on how the flurry of new magnetic
oxides could be useful for next-generation spintronics devices.

The final version of this review article has been published in IEEE Trans. Electron. Devices, 54, 1003 (2007)

\end{abstract}


%
\IEEEpeerreviewmaketitle

\section{Introduction}
%
%
%
%
\PARstart{I}{n} conventional electronics, information is encoded by the electron charge. In spintronics, the
electron spin is used as an additional degree of freedom to perform logic operations, store information, etc.
Spintronics exploits the spin-dependent electronic properties of magnetic materials and semiconductors
\cite{zutic2004}. The first widely studied spintronics effect was giant magnetoresistance (GMR), discovered in
metallic multilayers (e.g. Fe/Cr or Co/Cu superlattices) in the 1980's \cite{baibich88}. Since then, a large
variety of spintronics effects have been observed and studied \cite{zutic2004}, some of which are reviewed in
other articles of this Special Issue.

Another spintronics effect that has given rise to a large number of experimental and theoretical studies is
tunnel magnetoresistance (TMR). As we will show in the next part, large TMR effects can be obtained if special
magnetic materials (the so-called half-metals), having a finite density of states at the Fermi level for one
spin-direction and a gap for the other spin-direction, are used. Many materials that were conjectured to present
this unusual electronic structure were magnetic oxides. This opportunity of strongly enhancing spintronics
effects using magnetic oxides bridged a gap between the metal spintronics community and that of transition-metal
oxides. This latter family of researchers was mostly focusing on the recently discovered
high-critical-temperature superconductors \cite{bednorz86} and their growth in thin films using techniques like
pulsed laser deposition. This experience proved crucial for the fabrication and understanding of spintronics
structures based on magnetic oxides.

Since the first pioneering TMR results obtained on tunnel junctions based on manganese perovskite oxides
\cite{lu96}, the interest for oxides in spintronics has increased at a quick pace. This research was boosted by
the discovery of ferromagnetism in diluted magnetic semiconductors like (Ga,Mn)As \cite{ohno96} which triggered
an intense activity on oxide semiconductors (ZnO, TiO$_2$, etc) doped with magnetic ions. More recently, the
renewed interest in multiferroic materials and their growth in thin film form has provided novel opportunities
for oxides in spintronics.

In this paper, we will review some of the most significant achievements using oxides in spintronics experiments.
Even more than in the case of spintronics with 3d metals or semiconductors like GaAs, oxide spintronics is
complicated by materials science (growth, characterization, materials physics) aspects that have to be properly
addressed in order to perform clean and understandable spintronics experiments. In each part of this review, a
summary of the experimental status for materials issues will be provided, before getting into the details of
some remarkable spintronics results. Finally, prospects on possible future spintronics devices using magnetic
oxides will ge given.

\subsection{Tunnel magnetoresistance}

Before surveying recent results on oxide spintronics, let us recall briefly the physics of tunnel
magnetoresistance. Although the tunnel magnetoresistance (TMR) in magnetic tunnel junctions (MTJs) is known
since the experiments of Julli\`ere \cite{julliere75} thirty years ago, this phenomena has only been under
strong focus during the last ten years. This activity, pushed by the possible use of TMR-based devices such as
non volatile magnetic random access memories (MRAM), has followed the observation of a large TMR effect (16\%)
at room temperature in CoFe/Al$_2$O$_3$/Co junctions \cite{moodera95}. The last decade has led not only to
improvements in the amplitude of the TMR effect but also to advances in its understanding. Several recent
reviews report on the TMR effect \cite{tsymbal2003,zutic2004,fert2006}. In the following we only emphasize some
significant aspects.

A MTJ is composed of two ferromagnetic metallic electrodes sandwiching a very thin insulating barrier the
carriers have to cross by quantum-mechanical tunneling. The TMR is defined as the variation of resistance
between the parallel (R$_p$) and the antiparallel (R$_{ap}$) state of magnetizations of the two magnetic
electrodes:

\begin{equation}
TMR=\frac{R_{ap}-R_p}{R_p} \label{tmrdef}
\end{equation}

In the early stage of research on TMR, following the approach developed by Meservey and Tedrow
\cite{tedrow70,tedrow71}, this TMR effect has been related to the spin polarization P$_1$ and P$_2$ of the two
electrodes by the Julli\`ere formula \cite{julliere75}:

\begin{equation}
TMR=\frac{2 P_1 P_2}{1-P_1 P_2} \label{julliere}
\end{equation}

\noindent with P$_i$ (i=1,2) defined by the normalized difference between the density of states at the Fermi
level for the majority (N$_{i\uparrow}(E_F)$) and minority (N$_{i\downarrow}(E_F)$) spin electrons, i.e.

\begin{equation}
P_i=\frac{N_{i\uparrow}(E_F)-N_{i\downarrow}(E_F)}{N_{i\uparrow}(E_F)+N_{i\downarrow}(E_F)}
\end{equation}

Further experiments have shown that the spin polarization determined from tunneling experiments is not an
intrinsic property of the ferromagnetic electrodes but depends on the barrier \cite{deteresa99a,deteresa99},
i.e. the polarization represents an effective spin polarization of the tunneling probability. Recent theoretical
work has clearly put forward the role of interfacial bonding and the importance of symmetry in determining the
hybridization between the Bloch states of the metal and the slowly decaying states in the insulator
\cite{tsymbal2000,oleinik2000,oleinik2002,mavropoulos2000,maclaren99,butler2001}. Depending on the barrier
material, symmetry rules can select different wave functions for each spin direction and lead to different signs
of the spin polarization for the same ferromagnetic electrode. This effect is even reinforced by the record
magnetoresistance values that have been obtained recently with epitaxial MgO barriers
\cite{faure-vincent2003,yuasa2004,yuasa2004a,parkin2004,yuasa2006} following theoretical predictions
\cite{maclaren99,butler2001}.

\subsection{Why oxides ?}

Another way to obtain very large TMR ratio is to use half metallic electrodes. It is indeed very straightforward
from the Julli\`ere formula that materials with only one spin direction at the Fermi level (i.e. a total spin
polarization) should produce record TMR. Among the materials that have been predicted to be half-metals, many
are oxides (for a description and classification of half-metals, see \cite{coey2002,coey2004}). This triggered
an important activity on magnetic oxide thin films, with the main objective of making oxide tunnel junctions and
measure TMR effects. This will be reported in Part \ref{partII} as well as results concerning the search for
novel materials with a large spin polarization and a high T$_C$.

More recently, it was realized that the large number of degrees of freedom existing in transition metal oxides
could be used to design their physical properties according to some specific required function. In addition, the
tendency towards device miniaturization has emphasized the need for multifunctional materials, i.e. materials
that can perform more than one task or that can be manipulated by several independent stimuli. The large variety
and the tunability of the physical properties exhibited by transition-metal oxides (ferroelectricity,
ferromagnetism, antiferromagnetism, metallicity, superconductivity, optical properties, etc) like perovskites
thus provide tremendous advantages for spintronics by bringing additional functionalities that do not exist in
more conventionally used materials. In this perspective, research on oxide-based spintronics might well be only
at its beginning. Parts \ref{partIII} and \ref{partV} will present such oxide materials that are emerging in the
field of spintronics, namely diluted magnetic oxides and multiferroic materials. Part \ref{partIV} summarizes
the results obtained using the innovative concept of spin filtering using ferromagnetic or ferrimagnetic
insulating oxides as the tunnel barrier. Finally, Part \ref{partV} will discuss the potential of oxides for
prospective spintronics applications.

\section{Highly spin-polarized oxides}
\label{partII}

\subsection{Manganites}

Even though the term "half-metallic ferromagnet" was coined in 1983 by de Groot \emph{et al} for Mn-based
Heussler alloys \cite{degroot83}, predictions of a half-metallic character was extended to double-exchange
oxides such as Fe$_3$O$_4$ in 1984 \cite{yanase84}, CrO$_2$ in 1986 \cite{schwartz86} and manganites in 1996
\cite{pickett96}. The half-metallic character of manganites was demonstrated for the first time by spin-resolved
photoemission experiments \cite{park98b}, through which a positive spin-polarization in excess of 90 \% was
determined. By that time, manganite-based magnetic tunnel junctions with large TMR values had already been
fabricated and measured \cite{lu96,sun96,viret97} but it was only some years after Park \emph{et al}'s
publication \cite{park98b} that TMR measurements corresponding to spin-polarizations exceeding this value were
measured \cite{bowen2003}.

Manganites crystallize in the simple perovskite structure. The parent compound LaMnO$_3$ is an antiferromagnetic
charge-transfer insulator. Substituting a fraction of the La$^{3+}$ ions by divalent ions such as Sr$^{2+}$
induces a transition to a ferromagnetic and metallic state for substitution levels of about 17 \%
\cite{imada98}. Mixed-valence manganites such as La$_{1-x}$Sr$_x$MnO$_3$ are double-exchange ferromagnets with a
maximum Curie temperature of $\sim$360K for x=0.30-0.40. As the literature of manganites is immense and has been
extensively reviewed many times, we will refer the reader to such articles
\cite{imada98,coey99,salamon2001,ziese2002,haghiri2003,dorr2006} and focus in the following on the use of
manganites in magnetic tunnel junctions

\subsubsection{TMR}

The first TMR measurement on magnetic tunnel junctions with manganite electrodes was reported by the group of
Sun (IBM Yorktown Heights) in 1996 \cite{lu96,sun96}. The best results were obtained on junctions using
optimal-doped La$_{2/3}$Sr$_{1/3}$MnO$_3$ (LSMO) electrodes and 3 to 6 nm thick SrTiO$_3$ (STO) barriers. The
junctions were defined by a combination of optical lithography and ion-beam etching. A maximum TMR of 83 \% was
found \cite{lu96} (at 4.2K) which, according to Julli\`ere formula \cite{julliere75}, corresponds to a
spin-polarization of 54 \% for the LSMO electrodes. Lu \emph{et al}'s pioneering paper also reported several
hallmark features of manganite MTJs, such as the early disappearance of TMR upon increasing temperature and the
typical deviation from the parabolic behavior found in conductance (G) curves at low bias voltage. A few months
after these first two papers, the Sun group reported a TMR of $\sim$ 400 \%, corresponding to P$\simeq$81\%
\cite{sun97}. This was soon followed by Viret \emph{et al}'s paper that independently reported a 450 \% TMR at
4.2K in LSMO/STO/LSMO junctions \cite{viret97}.

\begin{figure}
\centering
\includegraphics[width=\columnwidth]{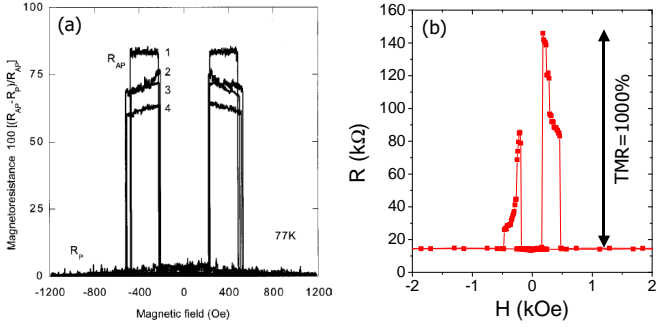}
\caption{(a) R(H) curves collected on four LCMO/NdGaO$_3$/LCMO junctions, corresponding to a maximum TMR of 630
\% (with the definition of Eq. \ref{tmrdef}), measured at 77K (reproduced from reference \cite{jo2000}); the
different curves correspond to different junctions. (b) TMR of $\sim$ 1000 \% at 4.2K in a LSMO/LAO/LSMO
junction (LAO is for LaAlO$_3$) \cite{bibes2006}.} \label{tmr_lsmo}
\end{figure}

Subsequent publications by the Sun group and others reported increasingly large TMR values (see figure
\ref{tmr_lsmo}), up to a TMR of 1850 \% in a LSMO/STO/LSMO MTJ, as found by Bowen \emph{et al} in 2003
\cite{bowen2003}. This record TMR corresponds to a spin-polarization of 95 \%, i.e. a virtually half-metallic
character for LSMO.

\subsubsection{Temperature dependence of the TMR}

Soon after the first observations of these large TMR ratios, much attention was brought to the temperature
dependence of the TMR. Indeed, in manganite tunnel junctions, the TMR decreased rather rapidly with temperature
and disappeared at a critical temperature T* (typically 200K in early reports), well below the Curie temperature
of the electrodes (maximized at 360K in LSMO). This problem motivated both theoretical \cite{lyu99} and
experimental research on manganite-based junctions and also stimulated the search for other half-metallic
ferromagnets with higher T$_C$.

Several explanations have been invoked to explain the difference between T* and T$_C$: defects in the tunnel
barrier causing spin-flips \cite{obata99,odonnell2000a}, non-optimal magnetic properties at manganite/barrier
interfaces (either due to oxygen deficiency \cite{viret97}, to phase-separation \cite{jo2000}, etc). We note
that the properties of manganite interfaces have been explored extensively through the study of manganite
surfaces \cite{park98a} and manganite/insulator interfaces, in thin films \cite{sun99,bibes2001e} and
heterostructures \cite{jo99,izumi2001}. Disrupted magnetic and electronic properties (the main features of which
are a lower T$_C$, a lower saturation magnetization, a larger resistivity and a larger low-temperature
magnetoresistance) are consistently observed in manganite ultrathin films. It has been suggested that this
disruption is due to strain \cite{millis98b}, charge-transfer \cite{yamada2004} or atomic-scale disorder
promoting charge trapping \cite{bibes2002}.

\begin{figure}
\centering
\includegraphics[width=\columnwidth]{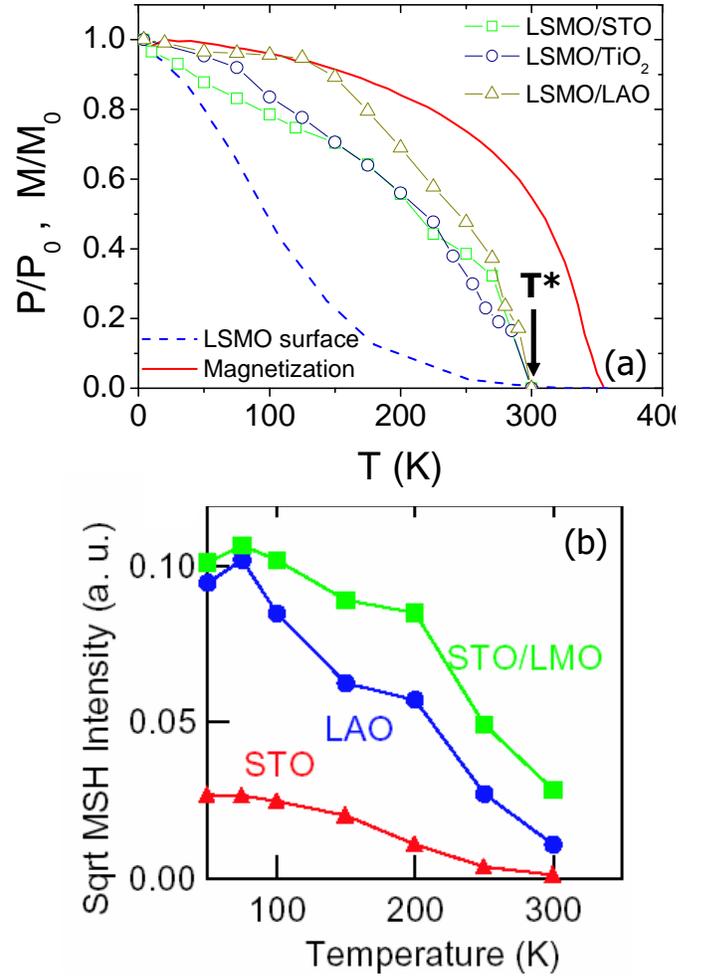}
\caption{(a) Temperature dependence of the spin-polarization of several manganite interfaces, as measured by
Garcia \emph{et al} \cite{garcia2004}. The graph also includes the temperature dependence of the magnetization
of a LSMO film, and that of a LSMO free surface (as reported in \cite{park98a}). (b) Temperature dependence of
the magnetization of several manganite interfaces measured by magneto-optical second harmonic generation (MSH)
by Yamada \emph{et al} \cite{yamada2004}. LMO stands for LaMnO$_3$ and the STO/LMO curve corresponds to a
LSMO/LMO(0.8 nm)/STO engineered interface.} \label{tdep_lsmo}
\end{figure}

Recently, systematic studies of interface effects in manganite heterostructures and manganite tunnel junctions
have addressed this issue in greater detail. Garcia \emph{et al} measured the temperature dependence of the TMR
in manganite tunnel junctions with either STO, LAO or TiO$_2$ barriers \cite{garcia2004}. Remarkably, for all
these junctions, T* is virtually the same and close to 300K, irrespective of the barrier material and thus of
the type of manganite/barrier interface. The spin-polarization actually does not decay very rapidly with
temperature (see figure \ref{tdep_lsmo}a), as opposed to the spin-polarization of a manganite free surface
\cite{park98a}, but follows a Bloch law

\begin{equation}
P(T)=P_0 (1-\alpha T^{3/2})
\label{bloch_law}
\end{equation}

\noindent with $\alpha$ values only slightly larger than the one found for bulk LSMO (from M(T) data)
\cite{garcia2004}. Recent magneto-optical data on LSMO/LAO and LSMO/STO interfaces indicate that the interface
magnetization vanishes close to 300K for both interface types \cite{yamada2004} (see figure \ref{tdep_lsmo}b),
in good agreement with Garcia \emph{et al}'s results.

Figure \ref{tdep_lsmo}b also includes the M(T) for a LSMO/STO interface in which 2 unit cells of LaMnO$_3$ have
been inserted been the LSMO and the STO layers. The purpose of the LMO layer is to compensate for charge
transfer that was found to occur at the LSMO/STO interface \cite{kumigashira2006}, increasing the hole density
in the last LSMO unit-cells. Even though better magnetic properties are obtained for this engineered interface,
the efficiency of this approach to enhance the TMR and its temperature stability remains to be demonstrated
clearly \cite{ishii2006}.

\subsubsection{Bias dependence of the TMR}

\begin{figure}
\centering
\includegraphics[width=\columnwidth]{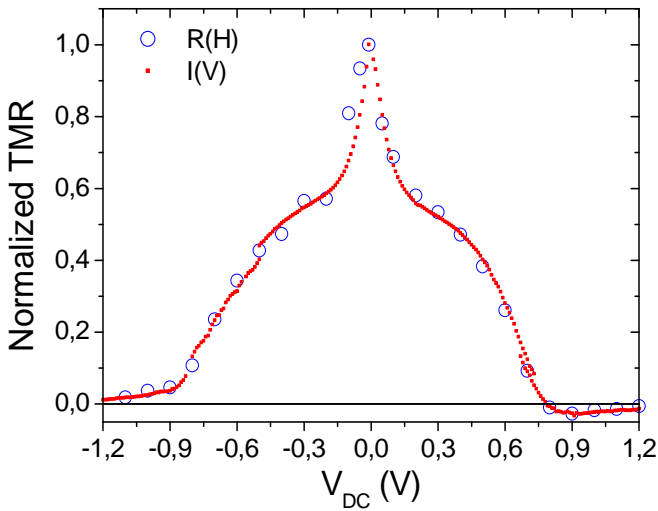}
\caption{Bias dependence of the TMR (normalized to its value at 10 mV) in a LSMO/STO(2.8nm)/LSMO junction, at
4.2K \cite{bowen2005}.} \label{tmr_v_lsmo}
\end{figure}

In addition to its temperature dependence, the bias-dependence of the TMR in manganite MTJs has also received
some attention. One of the main features of the TMR(V) in manganite-based junctions is the rapid decrease of the
TMR with bias voltage, at least in the low ($\leq 0.2$V) bias range \cite{sun97,sun98,bowen2005}. We note that
this behavior is accompanied by a specific zero-bias anomaly (ZBA) in the conductance curves. In magnetic tunnel
junctions, a ZBA is usually observed and ascribed to the absorption/generation of spin-waves by tunneling
electrons \cite{zhang2001}. The ZBA in manganite junctions has been studied theoretically by Gu \emph{et al}
\cite{gu2001} who proposed that due to the double-exchange interaction, the low-bias conductance is proportional
to $\mid V \mid ^{3/2}$, in agreement with experimental results \cite{lu96,bowen2005,bowen2003,bibes2006a}. Such
magnon excitations depolarize the spin of the carriers hence decreasing the effective spin-polarization and thus
the TMR. At larger bias, Bowen \emph{et al} have reported the observation of a plateau in the TMR, close to
V=0.35 V, followed by another strong decrease beyond $\sim$400 mV \cite{bowen2005}, see figure \ref{tmr_v_lsmo}.
This inflection point corresponds to the onset of tunneling into the spin-down conduction band of LSMO (as
predicted by Bratkovsky \cite{bratkovsky97}) whose position in energy above the Fermi level can thus be
precisely inferred from TMR measurements. The value found (380 meV) is in good agreement with results from
spin-polarized inverse photoemission \cite{bertacco2002}. We note that in MTJs based on half-metallic Heussler
alloys, an analogous influence of the spin-dependent density of states on the TMR(V) has been reported recently
\cite{sakuraba2006}. For further details on the bias dependence of the TMR in LSMO MTJs and the physics of
coherent tunneling in such structures, the reader is referred to Bowen \emph{et al} \cite{bowen2007}.

\subsubsection{Fundamental tunneling studies with manganite-based junctions}

Given their almost total spin-polarization, manganites are particularly well suited for performing fundamental
spin-polarized tunneling studies. For instance, they can be used to probe the spin-dependent density of states
of a material used as an electrode, as in the previous example. One can replace the manganite counter-electrode
by some other ferromagnetic metal and get some insight on the electronic properties of this metal, as well as on
those of the barrier material, via the analysis of the junction conductance curves and TMR(V) data. This is
exemplified by the results of de Teresa \emph{et al} \cite{deteresa99a,deteresa99} on LSMO/STO/Co junctions. In
such MTJs, the TMR at low bias is negative (i.e. the spin-polarization of Co at the interface with STO is
negative) and shows a negative maximum at about V=0.4V for electrons tunneling from LSMO to Co. Upon further
increasing the voltage, the TMR changes sign to become positive at V$\simeq$-0.8V. These results were
qualitatively confirmed by Hayakawa \emph{et al} in Co$_{90}$Fe$_{10}$/STO/LSMO junctions
\cite{hayakawa2002,hayakawa2002a}. It was argued that this very peculiar dependence arises from specific bonding
effects at the STO/Co interface, favoring the tunnel transmission of Co d states that have a negative
spin-polarization at E$_F$. Theoretical calculations also predicted the appearance of a sizeable magnetic moment
on the Ti ions adjacent to the interface, antiparallel to the Co magnetization \cite{oleinik2002}. However,
recent experiments have been unable to confirm this prediction \cite{bowen2006}.

Experimentally, a negative spin-polarization for Co at the interface with epitaxial TiO$_2$ \cite{bibes2003a}
and LaAlO$_3$ \cite{garcia2005} barriers was also reported. The TMR(V) of Co/LAO/LSMO junctions is strikingly
similar to that of Co/STO/LSMO junctions. This strongly suggests that wave function symmetry selection by the
epitaxial barrier is a key parameter in determining the tunnel current in these junctions. Unfortunately, the
structure of perovskite insulating oxides that are traditionally used as tunnel barriers (SrTiO$_3$, LaAlO$_3$,
etc) is more complicated than that of rock-salt MgO and this may explain why so few complex electronic structure
calculations are available for these oxides compared to the MgO case
\cite{butler2001,mathon2001,zhang2004,belashchenko2005}. So far, only the complex band structure of SrTiO$_3$
has been calculated and published \cite{velev2005,bowen2006a}, even though preliminary results exist concerning
LaAlO$_3$ \cite{velev2006}, indicating a negative spin-polarization for a Co/LAO interface, in agreement with
experiments \cite{garcia2005}.

\subsection{Double perovskites}

Even though the study of manganite-based MTJs has deepened our understanding of spin-polarized tunneling and of
the interface properties of these complex oxides, the initial hopes of using them for room-temperature
spintronics applications have not been fulfilled. Devices exploiting some of the transport properties of
manganites at or close to room temperature have however been proposed, such as contactless potentiometers based
on the colossal magnetoresistance effect \cite{balcells2000} or bolometers \cite{goyal97,yang2006} but none of
these are strictly speaking spintronics devices.

On the other hand, the remarkable low-temperature spintronics properties of manganites soon motivated the search
for new half-metals with higher Curie points. Several high-T$_C$ compounds had been predicted to be
half-metallic in the 1980's, like Fe$_3$O$_4$ \cite{yanase84} and Heussler alloys \cite{degroot83} but the first
spin-polarization measurements on these systems with a complex structure were disappointing \cite{bona85}. Much
effort for the discovery of new high-T$_C$ half-metals focused on perovskites for which a great experience had
been accumulated through the study of manganite films and heterostructures.

The family of double-perovskites first received attention in the 1960's
\cite{patterson63,galasso66,nakagawa68,sleight62} but it is only in 1998 that it was first conjectured that
Sr$_2$FeMoO$_6$ (SFMO) should be half-metallic \cite{kobayashi98}. This compound had indeed some of the required
properties of a good half-metal, such as an almost integer magnetic moment and, remarkably a T$_C$ of 420K and
an intergrain magnetoresistance slowly decaying with temperature (which was thought to reflect better interface
properties than those of manganites \cite{hwang96}).

Most research on double-perovskites such as SFMO has focused on bulk samples and aimed at understanding better
their structural, electronic and magnetic properties, and the connection between them. The magnetic interaction
is a special type of double exchange in which the metallic band is formed by the overlap of Fe t$_{2g}$ states,
O 2p states and Mo t$_{2g}$ states \cite{sarma2000a}. The spin of the itinerant electrons is antiparallel to the
local moment of the Fe ions and parallel to the small moment carried by the Mo ions, and thus the
spin-polarization is expected to be negative. A first evidence for this mechanism was provided by Besse \emph{et
al} \cite{besse2002}. Within the double-exchange picture, the Curie temperature depends on band filling and
attempts to increase the T$_C$ of SFMO by La doping at the Sr sites have been successful \cite{navarro2001}.
Unfortunately the magnetization of these doped systems decreases with doping as the introduction of dopant ions
enhances the Fe/Mo disorder \cite{navarro2001,navarro2003}. Nevertheless, a few undoped double-perovskites have
T$_C$ above 500K such as Ca$_2$FeReO$_6$ \cite{granado2002,kato2004} and Sr$_2$CrReO$_6$ \cite{kato2002}.
Ca$_2$FeReO$_6$ has an insulating behaviour but Sr$_2$CrReO$_6$ is metallic \cite{kato2004}. The possible
half-metallic character of this latter compound is still debated \cite{vaitheeswaran2005,tang2005}.

The advantages of SFMO and other metallic double-perovskites are thwarted by some severe problems. The first one
is that optimal magnetic properties require a perfect three-dimensional ordering of the magnetic ions at the
perovskite B sites, Fe and Mo in the case of SFMO. When cationic disorder is present, the magnetization
decreases systematically due to antiferromagnetic coupling between Fe neighbors at antiphase boundaries
\cite{balcells2001,navarro2001a}. Disorder is also detrimental to the spin-polarization \cite{sarma2000}.
Another one is its strong reactivity to air and water, which makes processing and storage problematic
\cite{navarro2003a}.

Despite these difficulties, the growth of double-perovskite thin films was carried out in several laboratories.
Different growth techniques have been used \cite{rager2002,asano2001}, the more popular being pulsed laser
deposition (PLD) \cite{manako99,venimadhav2004,sanchez2004,fix2005,wang2006}. To obtain single-phase SFMO films
by PLD is very difficult. The growth pressure must be kept very low because at higher pressure the more stable
SrMoO$_4$ compound forms \cite{santiso2002,manako99}. Conversely, at such low pressures, metallic Fe and
Fe$_3$O$_4$ tend to appear \cite{santiso2002,besse2002a}, and thus yield overestimated magnetization values. In
addition, high growth temperatures (typically above 850$^\circ$C) are required to obtain a good Fe/Mo ordering.

\begin{figure}
\centering
\includegraphics[width=\columnwidth]{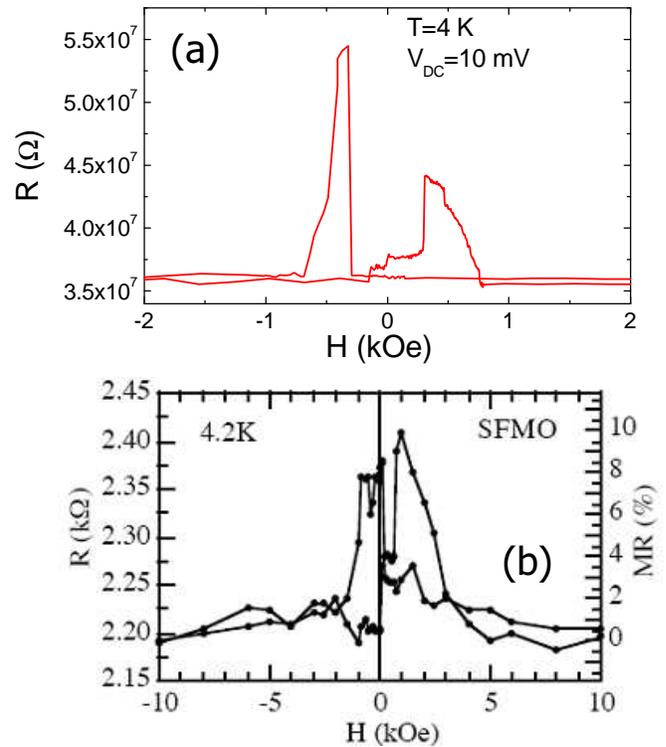}
\caption{TMR curves measured at 4.2K on a SFMO/STO/Co tunnel junction \cite{bibes2003} (a) and a SFMO/I/Co
tunnel junction \cite{asano2005} (b). In this last structure I is the native oxide appearing at the surface of
SFMO layers.} \label{sfmo}
\end{figure}

Several groups have attempted to measure the spin-polarization of SFMO, mostly in thin films. Values of 60 to 75
\% have been estimated through point-contact spectroscopy measurements \cite{auth2003,bugoslavsky2005}. Given
the poor compatibility of SFMO with typical perovskite barrier materials (like SrTiO$_3$) and its reactivity to
water, SFMO-based tunnel junctions are extremely challenging to fabricate and specific lithography processes
need to be developed. Using a special nanoindentation-based lithography technique \cite{bouzehouane2003}, it has
been possible to define Co/STO/SFMO nanometric ($\sim$ 20 nm in diameter) tunnel junctions and to measure their
TMR at low temperature, see figure \ref{sfmo}a. The TMR of such junctions is positive and ranges from 10 to 50
\% \cite{bibes2003}. As Co/STO/LSMO junctions give a TMR of up to -50 \% \cite{deteresa99}, the observation of a
TMR of +50 \% in Co/STO/SFMO indicates that the spin-polarization of SFMO is comparable to that of LSMO, but
with an opposite sign, i.e. at least -80 \%. More recently Asano \emph{et al} have also reported the observation
of a positive TMR of 10 \% at 4K in SFMO-based junctions, using a native oxide as the tunnel barrier and Co as
the top electrode \cite{asano2005} (see figure \ref{sfmo}b).

Other double-perovskite compounds have also been grown in thin films, especially Sr$_2$CrWO$_6$
\cite{philipp2003,venimadhav2006} and more recently Sr$_2$CrReO$_6$ \cite{asano2004} that has a T$_C$ of 635K
\cite{kato2002}. No data concerning the spin-polarization of these compounds is yet available.

Finally, we would like to point out that the physics of double perovskites have been recently reviewed by
Serrate \emph{et al} \cite{serrate2007}.

\subsection{CrO$_2$}

Chromium dioxide (CrO$_2$) is a metallic binary oxide crystallizing in the tetragonal rutile structure. The
moments of the Cr ions order ferromagnetically below T$_C$$\simeq$395K. At low temperature, the saturation
magnetic moment is 2 $\mu_B$/f.u., as expected for a collinear ordering of the Cr$^{4+}$ ions with a t$_{2g}^2$
electronic configuration. The Fermi level lies in the half-full d$_{yz}\pm$d$_{zx}$ band and CrO$_2$ is a
prototypical double-exchange system. The low-temperature resistivity is on the order of 1 $\mu\Omega$.cm in high
quality thin films \cite{watts2000}. More details on the properties of CrO$_2$ can be found in reference
\cite{coey2002}.

CrO$_2$ films can be grown by chemical vapor deposition (CVD) \cite{li99,ivanov2001}, thermal decomposition
\cite{ranno97} or pulsed laser deposition (PLD) \cite{shima2002}. A large spin-polarization can be expected for
CrO$_2$ from its electronic structure and the large powder magnetoresistance observed in polycrystalline samples
\cite{coey98a}, ascribed to spin-dependent tunneling between grains. Several techniques have been used to
measure directly the spin-polarization. Andreev reflection experiments with CrO$_2$-superconductor point
contacts \cite{soulen98,desisto2000,anguelouch2001} have indicate a very large P, up to 97 \%, at low
temperature. Spin-dependent tunneling experiments using a superconducting Al spin-detecting electrode
(Meservey-Tedrow technique) confirmed this result \cite{parker2002}. Despite these record spin-polarization
values, rather small TMR effects have been reported in CrO$_2$-based MTJs. Up to now, three groups have
published results on the fabrication and characterization of such MTJs, using CrO$_2$ as one electrode and Co as
the other \cite{barry2000,gupta2001,parker2004}. The selection of the barrier material is a problem as a native
oxide (Cr$_2$O$_3$) nanometric layer forms at the surface of the CrO$_2$ films. This layer can nevertheless be
used as a tunnel barrier, yielding a maximum TMR of -8\% \cite{gupta2001}. We note that Cr$_2$O$_3$ is a
multiferroic material \cite{astrov60}. Depositing Al to form a composite CrO$_x$-AlO$_x$ barrier has allowed to
increase this value to -24 \% \cite{parker2004}. These TMR values are far too small in view of the almost total
spin-polarization measured in Andreev reflection and Meservey-Tedrow experiments, likely due a modified
electronic structure of CrO$_2$ at the interface with the native barrier. The presence of impurity states inside
the barrier, making inelastic tunneling processes dominant, is also a likely explanation as pointed out by
Parker \emph{et al} \cite{parker2004}.

\subsection{Fe$_3$O$_4$}

Magnetite (Fe$_3$O$_4$) is the oldest known magnetic material. It is a ferrimagnet with a critical temperature
of T$_C$=858K. Fe$_3$O$_4$ crystallizes in the spinel structure whose A sites are occupied by Fe$^{3+}$ ions and
B sites by a mixture of Fe$^{2+}$ and Fe$^{3+}$ ions. As in most spinel ferrites, the ferrimagnetic ordering
arises from the strong antiferromagnetic coupling between the A and B sublattices. Within the B sublattice, the
Fe ions are coupled ferromagnetically. Given the mixed Fe valence in this sublattice, double-exchange
interaction can take place, as was theoretically described by Loos and Novak \cite{loos2002}. A half-metallic
state with a total negative spin-polarization is expected, the conduction band being formed by the overlap of
spin-down Fe t$_{2g}$ states with O 2p states. This is what is indeed indicated by electronic structure
calculations \cite{yanase84,degroot86,zhang91,antonov2003}.

\begin{figure}[h!]
\centering
\includegraphics[width=\columnwidth]{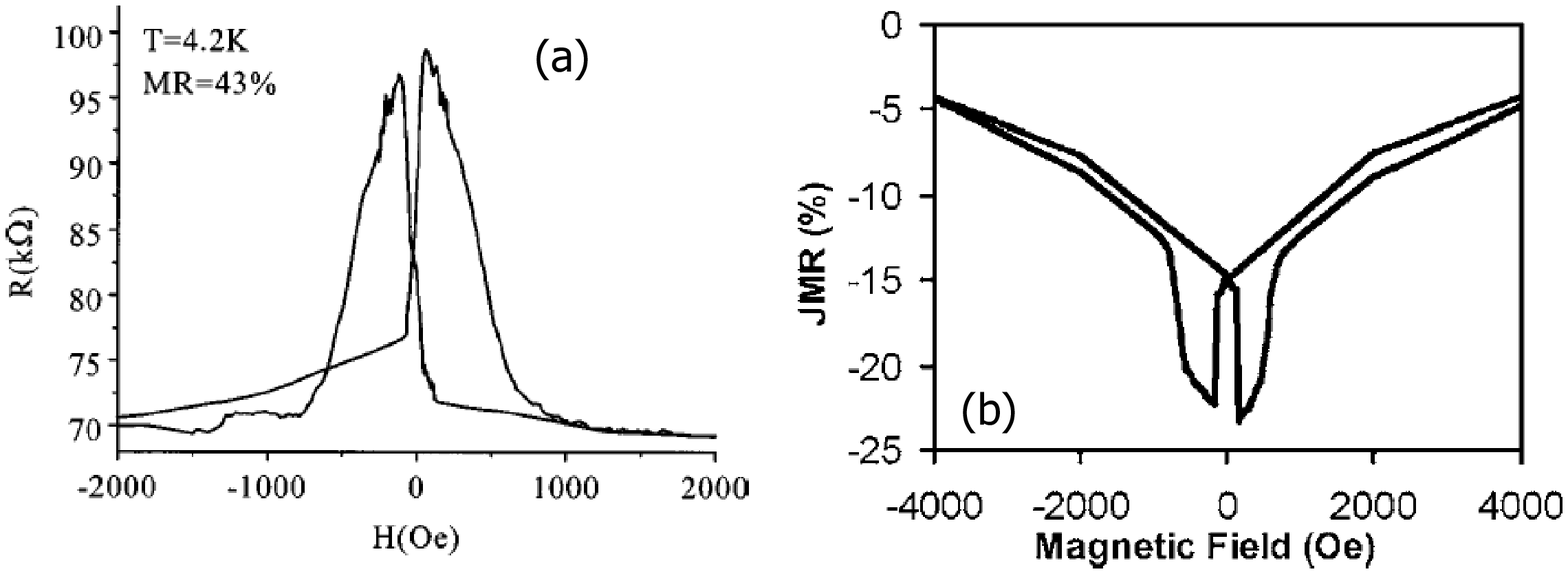}
\caption{(a) TMR curve measured at 4K in a Co/Al$_2$O$_3$/Fe$_{3-x}$O$_4$ junction (from \cite{seneor99}. (b)
TMR curve of a Fe$_3$O$_4$/MgTi$_2$O$_4$/LSMO junction, at 50K (from \cite{alldredge2006}).} \label{fe3o4}
\end{figure}

At room temperature, magnetite is a bad conductor, with a metallic behavior occurring in the $\sim$320-800K
range \cite{todo95}. Below room-temperature the resistivity is thermally activated, with a transition to a
charge-ordered insulating state below the Verwey transition temperature (T$_V\simeq$120K). The precise nature of
the electronic properties below T$_V$ is a long-standing problem in solid-state physics that is still strongly
debated as illustrated by the contradictory review articles published by Walz \cite{walz2002} and Garcia
\emph{et al} \cite{garcia2004a}.

The growth of magnetite thin films has been carried out by many groups, by a large number of deposition
techniques such as sputtering \cite{margulies96}, molecular-beam epitaxy \cite{voogt95} and pulsed laser
deposition \cite{gong97}. It is a consistent trend in the literature that the properties of magnetite thin films
deviate from those of the bulk. In particular, their magnetization is hard to saturate and often lower than the
bulk one \cite{margulies96,margulies97}. These effects are exacerbated by thickness reduction, and films as thin
as a few nm become superparamagnetic \cite{voogt98,eerenstein2003}. These disrupted magnetic properties are
accompanied by an increase in the electrical resistivity and a blurring or even disappearance of the Verwey
transition \cite{li98,eerenstein2002}. This has been convincingly related to the films microstructure and the
presence of antiphase boundaries \cite{eerenstein2002,eerenstein2003a}.

The spin-polarization of magnetite measured by spin-polarized photoemission spectroscopy is generally found to
be negative, ranging from -40 to -80 \% \cite{huang2002,huang2002a,dedkov2002,fonin2003,bataille2006}. This is
lower than what is expected from band structure calculations and several explanations have been put forward to
explain this discrepancy (surface reconstruction \cite{fonin2005}, surface defects, etc). The spin-polarization
values deduced from spin-dependent tunneling experiments show a large dispersion (see figure \ref{fe3o4}).
Seneor \emph{et al} \cite{seneor99} have found a positive spin-polarization of about 50 \% from TMR experiments
on Fe$_{3-x}$O$_4$/Al$_2$O$_3$/Co junctions (see figure \ref{fe3o4}a). This positive sign, also observed by
Aoshima \emph{et al} \cite{aoshima2003}, Yoon \emph{et al} \cite{yoon2004}, Bataille \emph{et al}
\cite{bataille2006a} and Reisinger \emph{et al} \cite{reisinger2004}, stands in contrast with the negative
spin-polarization calculated or measured by spin-polarized photoemission. A negative spin-polarization was also
found by the Suzuki group in LSMO/CoCr$_2$O$_4$/Fe$_3$O$_4$ \cite{hu2002}, LSMO/MgTi$_2$O$_4$/Fe$_3$O$_4$ (see
figure \ref{fe3o4}b) and LSMO/FeGa$_2$O$_4$/Fe$_3$O$_4$ junctions \cite{alldredge2006b}. Recently, Parkin
\emph{et al} have also measured a negative spin-polarization of -48 \% for Fe$_3$O$_4$/AlO$_x$ interfaces
\cite{parkin2006}. This large dispersion in the sign and values of the spin-polarization is likely to be related
to the strong sensitivity of the electronic properties of iron oxides to oxygen stoichiometry
\cite{tsymbal2000}.

A few other spintronics devices based on Fe$_3$O$_4$ have also been fabricated. This is the case of magnetic
tunnel transistors using a magnetite emitter, an alumina barrier and a Si collector \cite{yoon2004a}.
Current-in-plane (CIP) GMR structures with Au or Pt spacer layers have also been studied by van Dijken \emph{et
al} \cite{vandijken2004} and Snoeck \emph{et al} \cite{snoeck2006}. A maximum MR ratio of 5 \% was measured at
low temperature.

Efforts to integrate Fe$_3$O$_4$ into semiconductor structures \cite{kennedy99,lu2004,watts2004,reisinger2003}
have not permitted yet to fabricate devices that would exploit the spintronics properties of Fe$_3$O$_4$.
Surface and interface properties of Fe$_3$O$_4$ are still under investigation to understand spin-polarization
measurements and their connection with predictions from ab-initio calculations \cite{fonin2005,berdunov2004}.
Another issue is the role of antiphase boundaries \cite{eerenstein2002}. Defining devices with lateral sizes in
the deep sub 100 nm range might permit to avoid the contribution to antiphase boundaries and define better their
role on transport properties.

Finally, we note that Fe$_3$O$_4$ is a magnetoelectric material at low temperature, see for instance
\cite{miyamoto94,matsubara2005}. This potentiality for spintronics (discussed in part \ref{partV}) has not been
exploited yet.

\subsection{NiFe$_2$O$_4$}

Apart from Fe$_3$O$_4$, all other bulk spinel ferrites are insulating. This is the case of NiFe$_2$O$_4$
\cite{austin70} that has an inverse spinel structure in its bulk form and shows ferrimagnetic order below 850K
and an insulating character with a room temperature resistivity of $\sim$1k$\Omega$.cm \cite{brabers95}. Its
magnetic structure consists of two antiferromagnetically coupled sublattices. A first sublattice is formed by
ferromagnetically ordered Fe$^{3+}$ (3d$^5$, magnetic moment (M) : 5 $\mu_B$) ions occupying the tetragonal A
sites of the spinel AB$_2$O$_4$ structure, while the second sublattice contains ferromagnetically ordered
Ni$^{2+}$ (3d$^8$, M=2 $\mu_B$) and Fe$^{3+}$ (3d$^5$, M= 5$\mu_B$) ions occupying the octahedral B sites. This
type of ordering results in a saturation magnetization of 2 $\mu_B$/f.u. (f.u. : formula unit) or 300
emu.cm$^{-3}$. Recent electronic structure calculations have consistently estimated smaller gap values for
spin-down than for spin-up \cite{szotek2004,zuo2006,itoh2006}.

Remarkably, NiFe$_2$O$_4$ can be turned into a conductive material in thin films grown by sputtering on
SrTiO$_3$ substrates \cite{luders2005} in a pure Ar atmosphere. This conductive behavior (room temperature
resistivity: $\sim$ 100 m$ \Omega$.cm \cite{luders2006a}) is likely to be related to oxygen vacancies promoting
a mixed valence for the Fe ions \cite{luders2006c} as occurs in magnetite. In addition, these films exhibit
magnetic properties that radically differ from those of bulk NiFe$_2$O$_4$ \cite{mccurrie94}, namely a
saturation magnetization larger by up to $\sim$300 \% \cite{luders2005}. Preliminary X-ray magnetic circular
dichroism indicate that this enhanced magnetic moment is to a great extent due to cationic inversion between the
ions at the A and B site \cite{luders2006c}. Cationic inversion is known to occur in spinel ferrite thin films
\cite{venzke96,yang2005} and has also been reported in nanoparticles \cite{kim2002,chinnasamy2001,zhou2002}.

\begin{figure}
\centering
\includegraphics[width=\columnwidth]{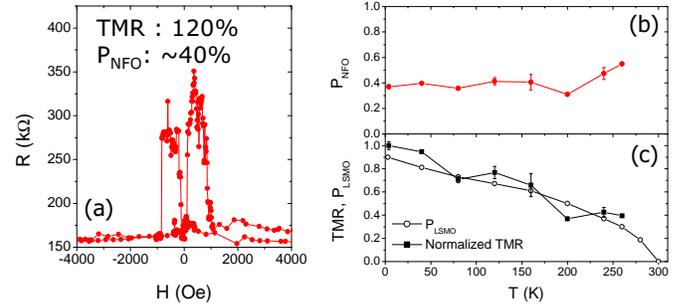}
\caption{(a) TMR curve measured at 4K for a magnetic tunnel junction with one LSMO and one conductive
NiFe$_2$O$_4$ electrodes (from \cite{luders2006a}). (b) Temperature dependence of the conductive NiFe$_2$O$_4$
spin-polalarization, as deduced from TMR vs temperature measurements (c) and the temperature dependence of the
spin-polarization of LSMO/STO interfaces (from \cite{luders2006b}).} \label{nfo}
\end{figure}

Such conductive NiFe$_2$O$_4$ films have been used as magnetic electrodes in NiFe$_2$O$_4$/SrTiO$_3$/LSMO
magnetic tunnel junctions, see figure \ref{nfo}. A TMR ranging in 15 to 140 \% has been measured at low
temperature \cite{luders2006b,luders2006a}, corresponding to a maximum spin-polarization of 45 \% for the
NiFe$_2$O$_4$ layer. Significantly, this value is close to the largest spin-polarization measured for magnetite
by tunneling experiments. In addition, as visible in figure \ref{nfo}b, this spin-polarization is virtually
constant up to 300K, in agreement with the high T$_C$ of NiFe$_2$O$_4$.

We note that recently, conductive (Mn,Fe)$_3$O$_4$ films have also been fabricated \cite{ishikawa2005}. Evidence
for a finite spin-polarization at room temperature has been provided by anomalous Hall effect
\cite{ishikawa2005} and preliminary TMR measurements in Co/AlO$_x$/(Mn,Fe)$_3$O$_4$ junctions \cite{tanaka2006}.

\subsection{SrRuO$_3$}

SrRuO$_3$ is a metallic ferromagnet (T$_C$=160K) crystallizing in the perovskite structure \cite{eom92}. Its
good electrical conductivity makes it a material of choice as electrode for ferroelectric capacitor
measurements. The conduction band is formed by the overlap of Ru t$_{2g}$ and O 2p orbitals. It is not a
double-exchange system but an itinerant ferromagnet \cite{allen96,mazin97,kiyama99}. Electronic structure
calculations have predicted a negatively spin-polarized state, with P$\simeq$-60 \% \cite{singh96}.

The first measurement of the spin-polarization of SrRuO$_3$ was performed by Worledge and Geballe using a
Meservey-Tedrow technique \cite{worledge2000}. A negative number was found, in agreement with theoretical
predictions, but with a much smaller value (-9\%). Subsequent Andreev reflection experiments reported a value of
50 \% \cite{nadgorny2003} (Andreev reflection is not sensitive to the sign of the spin-polarization). The
spin-polarization of SrRuO$_3$ has also been determined from TMR experiments in SrRuO$_3$/STO/LSMO tunnel
junctions \cite{takahashi2003,noh2004}. These two studies found a negative TMR, as expected from the Julli\`ere
model \cite{julliere75} and inferred a spin-polarization of about -10 \% for SrRuO$_3$, in agreement with
reference \cite{worledge2000}. It is interesting to note that the spin-polarization of SrRuO$_3$ decays roughly
like the magnetization \cite{takahashi2003}, i.e. it does not vanish prematurely like that of manganite does.
This observation confirms the robustness of SrRuO$_3$ to interface disorder, as inferred from magnetotransport
data through artificial grain-boundaries \cite{bibes99b}.

\section{Diluted magnetic oxides}
\label{partIII}

The search for novel magnetic materials, with ideally large spin-polarization, high T$_C$ and possibly
multifunctional characteristics has triggered an intense activity on doping non-magnetic semiconducting oxides
with magnetic ions. To a great extend the choice of oxide hosts was motivated by the prediction by Dietl
\emph{et al} of a T$_C$ above 300K in Mn-doped ZnO \cite{dietl2000}. This prediction opened a way to achieve
room-temperature operation with diluted magnetic semiconductors (DMS), which even nowadays is impossible with
(Ga,Mn)As, the prototypical DMS \cite{jungwirth2006}.

Here we will not review extensively the quickly growing field of diluted magnetic oxides (for more details see
references \cite{prellier2003,pearton2004,janisch2005,chambers2006}) but focus on some aspects most related to
their relevance for spintronics. The first important experimental report on diluted magnetic oxides was by
Matsumoto \emph{et al} \cite{matsumoto2001} who observed a ferromagnetic behavior at room temperature in
Co-doped (7\%) TiO$_2$. This was soon followed by papers reporting ferromagnetism in Co-doped \cite{ueda2001}
and V-doped ZnO \cite{saeki2001}. Later on, ferromagnetism was reported in several other diluted magnetic
oxides, with different oxide hosts (SnO$_2$, In$_2$O$_3$, HfO$_2$, Cu$_2$O, etc). It soon proved very difficult
to unambiguously demonstrate that the ferromagnetic behavior, typically observed using standard magnetometry
techniques (e.g. SQUID, AGFM, VSM), was intrinsic (e.g. due to some exchange mechanism resulting from the
substitution of some cation of the matrix by the magnetic dopant) rather than extrinsic (due to the formation of
parasitic ferro- or ferrimagnetic phases, in the form of nanometric clusters, filaments, etc). Actually, little
unambiguous evidence for intrinsic ferromagnetism in diluted magnetic oxides exists. The accumulation of
experimental results has also challenged theorists to imagine interaction mechanisms compatible with the data.
This has led to the development of novel concepts such as F-center exchange \cite{coey2005a} and d$^0$
ferromagnetism \cite{coey2005}, i.e. ferromagnetism without ions having partially filled d or f shells. Despite
these efforts, the physics of diluted magnetic oxides is not well understood and still the object of intense
experimental and theoretical activity.

In the following, we present some relevant results obtained in diluted magnetic oxides over the last few years.

\subsection{TiO$_2$}

Ferromagnetism in transition-metal-doped TiO$_2$ was first observed in anatase Ti$_{0.93}$Co$_{0.07}$O$_2$ films
grown by pulsed laser deposition by Matsumoto \emph{et al} \cite{matsumoto2001}. A few months later, Chambers
\emph{et al} also reported room temperature ferromagnetism in Co-doped anatase TiO$_2$ films grown by
oxygen-plasma-assisted molecular-beam epitaxy \cite{chambers2001,chambers2002}. A ferromagnetic behavior was
also found for sputtered Co-doped anatase TiO$_2$ films \cite{park2002}, and in Co-doped rutile TiO$_2$
\cite{matsumoto2001a}.

Controversy on the origin of ferromagnetism in this system appeared early after these first reports with the
observation of Co clusters by several groups \cite{kim2003,chambers2003,shinde2004}. These Co clusters can be
ferromagnetic or superparamagnetic and give rise to an anomalous Hall effect \cite{shinde2004}. Subsequent
reports have nevertheless provided strong indications that ferromagnetism is possible in cluster-free films in
which the Co dopants are homogeneously distributed within the material (see for instance \cite{griffin2006}). It
thus appears that only in a narrow range of growth conditions (i.e. dopant concentration, growth temperature,
growth pressure and growth rate) can cluster-free films be produced. This obviously makes comparisons between
films grown in different systems and \emph{a fortiori} with different growth techniques very difficult, if
possible at all.

The origin of ferromagnetism in doped TiO$_2$ has been the object of much theoretical
\cite{park2002a,janisch2006} and experimental efforts \cite{shinde2003,mamiya2006}. As before, it is hard to
summarize the whole literature as contradictory reports have been published. Nevertheless, the Kawasaki group
has provided indications for carrier-mediated ferromagnetism in Co-doped TiO$_2$ through systematic studies of
transport properties like the anomalous Hall effect \cite{toyosaki2004} and X-ray photoemission spectroscopy
experiments \cite{quilty2006}. In this system, carriers are usually provided by oxygen vacancies and their
concentration adjusted by the oxygen partial pressure during growth (see for instance \cite{jaffe2005}).
However, the role of oxygen vacancies is not restricted to electron doping as their presence was found to
influence the nucleation of metallic clustered phases \cite{kim2002a}. Furthermore, oxygen vacancies alone (i.e.
in the absence of magnetic dopants) could give rise to ferromagnetism. For instance, Yoon \emph{et al} recently
reported a ferromagnetic behavior in TiO$_{2-\delta}$ films \cite{yoon2006}. In addition, other types of defects
(e.g. structural defects) could also be relevant for ferromagnetism \cite{kaspar2005}.

\begin{figure}
\centering
\includegraphics[width=\columnwidth]{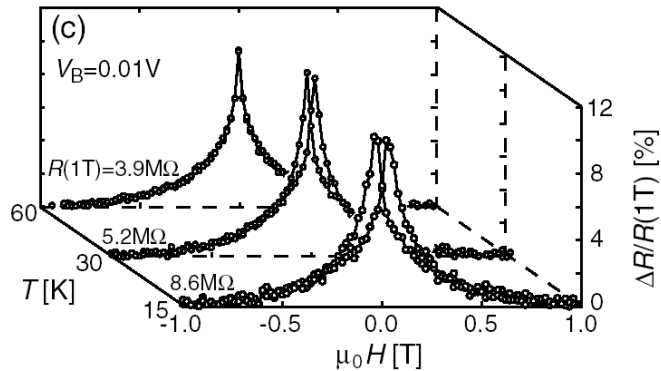}
\caption{Resistance vs field curves measured at different temperatures in a
Ti$_{0.95}$Co$_{0.05}$O$_{2-\delta}$/AlO$_x$/Co$_{0.9}$Fe$_{0.1}$ junction. (from \cite{toyosaki2005}).}
\label{cotio2}
\end{figure}

Even though the observation of anomalous Hall effect may be taken as evidence that charge carriers are
spin-polarized, a more direct proof is provided by spin-dependent tunneling experiments. Recently, Toyosaki
\emph{et al} reported the fabrication and characterization of
Co$_{90}$Fe$_{10}$/AlO$_x$/Ti$_{0.95}$Co$_{0.05}$O$_{2-\delta}$ magnetic tunnel junctions
\cite{toyosaki2005,toyosaki2006}. Hysteretic R(H) curves were observed at low temperature (see figure
\ref{cotio2}), with a maximum MR of 11 \%. The MR disappeared at about 200K, which was attributed to the
presence of defects in the barrier. These preliminary results remain to be confirmed by further experiments on
better-quality junctions.

Finally, we note that a reversible electric-field modulation of the magnetization of a
Ti$_{0.93}$Co$_{0.07}$O$_2$ film was reported in a
PbZr$_{0.2}$Ti$_{0.85}$O$_3$/Ti$_{0.93}$Co$_{0.07}$O$_2$/SrRuO$_3$ structure \cite{zhao2005}. The exploitation
of this effect in MTJs based on Co-doped TiO$_2$ could allow the design of novel multifunctional spintronics
devices.

\subsection{ZnO}

Ferromagnetism in transition-metal-doped ZnO was first reported by Ueda \emph{et al} in 2001 \cite{ueda2001}.
The measured magnetic moment was 2 $\mu_B$/Co in Zn$_{0.85}$Co$_{0.15}$O films. Another early report of
ferromagnetism in V-doped ZnO was also published by the same group \cite{saeki2001}. Since these first papers,
ZnO films doped with different 3d ions have been grown by a number of techniques, the most popular being pulsed
laser deposition. Reviews of the research status of doped ZnO films have been published recently
\cite{janisch2005,ozgur2005}. As with TiO$_2$, the precise origin of ferromagnetism is not clearly established
but several theoretical models have been put forward to describe the physics of transition-metal-doped ZnO. They
can be divided in two categories: (i) models based on Zener double-exchange or RKKY interactions
\cite{dietl2000,kittilstved2006a}; (ii) models based on F-center exchange \cite{coey2005a}. Reports of the
presence of ferromagnetic Co clusters \cite{maurice2006} or parasitic phases \cite{kundaliya2004} have
questioned the possibility of an intrinsic ferromagnetic behavior.

Nevertheless, some reports have evidenced a strong spd coupling, indicative of a coupling between the magnetic
dopant and carriers \cite{ando2001}. More recently, a detailed analysis of Co-doped ZnO and Mn-doped ZnO films
\cite{kittilstved2005} has shown that ferromagnetism is carrier-related. The introduction of p-type carriers by
N dopants induces ferromagnetism in Mn-doped ZnO, and n-type carriers provided by interstitial Zn are necessary
for ferromagnetic Co-doped ZnO \cite{kittilstved2006}. These experimental results are in agreement with
theoretical predictions \cite{dietl2000,wang2004,spaldin2004}.

The search for a finite spin-polarization of carriers has not been very successful but recently Xu \emph{et al}
\cite{xu2006} and Peng \emph{et al} \cite{peng2006} have measured anomalous Hall effect in Co-doped ZnO.
Magnetic tunnel junctions using a Co-doped ZnO electrode and giving rise to a magnetoresistance effect have been
measured by Rode \emph{et al} \cite{rode2006}. However, the influence of Co-clusters present in the ZnO films
has to be clearly studied before drawing conclusions on a possible intrinsic spin-polarization.

\subsection{(La,Sr)TiO$_3$}

When SrTiO$_3$ is doped with La, a transition from an insulating to a metallic state occurs for a La
concentration as low as $\sim$5 \% \cite{fujimori92,tokura93,sunstrom92}. La$_{1-x}$Sr$_x$TiO$_3$ is a
paramagnetic metal with a low-temperature resistivity of about 200 $\mu \Omega$.cm for the x=0.5 compound. The
temperature dependence of its resistivity is dominated by electron-electron scattering mechanisms, evidencing
the strong electronic correlations at play in this system. The importance of correlations become particularly
visible as x increases \cite{furukawa99} and the system is driven closer to the LaTiO$_3$ end compound that is a
prototypical antiferromagnetic Mott-insulator.

Epitaxial films of La$_{0.5}$Sr$_{0.5}$TiO$_3$ have been grown by a few groups \cite{wu2000,cho2001}, mostly
motivated by its possible use as electrode for ferroelectric capacitors or as a transparent conductor. The idea
of doping (La,Sr)TiO$_3$ (LSTO) films with magnetic ions was introduced by Zhao \cite{zhao2003} \emph{et al}. As
a host material, LSTO differs from most other systems by its strongly-correlated character and its large carrier
density. Its perovskite structure is an advantage for film growth and for combining it with other functional
oxides like manganites or multiferroics. A ferromagnetic behavior with a Curie temperature in the 500K range was
found by Zhao \emph{et al } \cite{zhao2003}, for a doping level of only 1.5 \%. The films resistivity was very
sensitive to oxygen pressure during growth, and a metallic state was found for pressures lower than
$\sim$10$^{-4}$ mbar, indicating a clear role of oxygen vacancies in transport. This was confirmed by annealing
studies \cite{qiao2004}. A ferromagnetic behavior was also found for Co-doped LSTO films by Ranchal \emph{et al}
\cite{ranchal2005}. Auger electron spectroscopy or high-resolution transmission electron microscopy with
electron energy loss analyses did not identify parasitic phases nor Co-rich clusters \cite{herranz2006}.
Preliminary X-ray absorption spectroscopy experiments indicate that the Co ions are in a 2+ state in a
octahedral environment \cite{petroff2006}.

\begin{figure}
\centering
\includegraphics[width=\columnwidth]{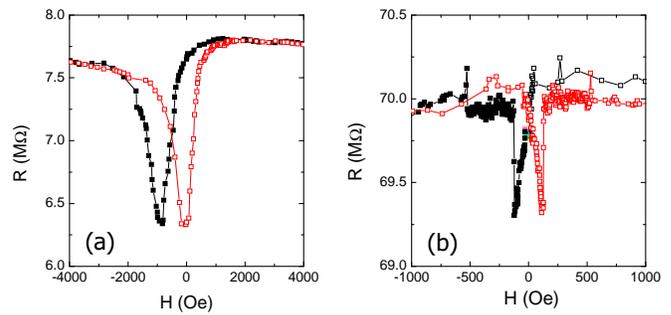}
\caption{(a) TMR curve measured at 4K and 10 mV on a Co-LSTO/LaAlO$_3$/Co/CoO tunnel junction (see reference
\cite{herranz2006}. (b) TMR curve measured at 4K and 10 mV on a Co-LSTO/SrTiO$_3$/Co nanojunction defined by the
process described in \cite{bouzehouane2003}.} \label{colsto}
\end{figure}

As a high T$_C$ ferromagnetic metal, Co-doped LSTO is an interesting material for spin-injection. To determine
its spin-polarization, Herranz \emph{et al} fabricated Co/LaAlO$_3$/Co-LSTO magnetic tunnel junctions and
measured their magnetotransport characteristics \cite{herranz2006}. A typical TMR curve obtained on such
junctions is represented in figure \ref{colsto}. The analysis of the bias dependence of the TMR in several
Co/LaAlO$_3$/Co-LSTO junctions in terms of defect-assisted tunneling yields a value of $\sim$-80 \% for the
spin-polarization of Co-LSTO. We note that a TMR was also observed on junctions with SrTiO$_3$, but with a lower
value \cite{herranz2006a}, see figure \ref{colsto}b. The TMR decreases with temperature and disappears around
200K, likely due to spin-depolarization inside the barrier or to the impossibility of achieving an antiparallel
state above this temperature.

Despite these encouraging tunneling results, the physics of Co-doped LSTO remains unclear. Recent studies of
transition-metal doped SrTiO$_3$ films, showing a paramagnetic behavior \cite{zhang2006}, suggest that the
presence of La and of hence a large carrier density is required for ferromagnetism. Many issues are still
unaddressed, like the role of oxygen vacancies or the dopant type. On this latter point, the studies of
(La,Sr)(Ti,Cr)O$_3$ bulk samples by Inaba \emph{et al} \cite{inaba2005} and Iwasawa \emph{et al}
\cite{iwasawa2006} might bring interesting insight.

\section{Spin-filtering}
\label{partIV}

The concept of spin filter relies on the use of a ferromagnetic or ferrimagnetic insulating tunnel barrier. In a
ferromagnetic and insulating material, the conduction bands are spin split by exchange leading to two different
barrier heights for spin-up and spin-down electrons. The combination of a non-magnetic electrode with a
ferromagnetic barrier yields two very different currents for spin-up and spin-down electrons due to the
exponential dependence of the transmission with the barrier height, resulting theoretically in a very large spin
polarization. Such a non-magnetic-metal/ferromagnetic-insulator bilayer constitutes a kind of artificial half
metal. Following early experiments by Esaki \emph{et al} \cite{esaki67}, the validation of the spin-filter
concept and the determination of its efficiency in spin-polarizing the current was reported by Moodera's group
\cite{moodera88}. In these first experiments the spin polarization of the current was measured by a
superconducting counter electrode at very low temperature. The spin polarisation can also be checked by a
metallic ferromagnetic counter-electrode which also acts as a spin analyser \cite{leclair2002}.

Several ferromagnetic or ferrimagnetic oxides, EuO \cite{santos2004}, NiFe$_2$O$_4$
\cite{luders2006,luders2006a}, CoFe$_2$O$_4$ \cite{chapline2006}, BiMnO$_3$ \cite{gajek2005} and
La$_{0.1}$Bi$_{0.9}$MnO$_3$ \cite{gajek2006a,gajek2006} have been used as magnetic insulating barriers in such
spin filters.

\subsection{EuO}

The efficiency in spin polarizing the current by an EuO barrier has been measured using the Meservey and Tedrow
technique in Al/EuO/Ag and Al/EuO/Y/Al structures \cite{santos2004}. A spin polarisation of 29\% at low
temperature has been measured in samples where a thin Y layer (5nm) is introduced at the top interface in order
to obtain a purer EuO phase.

\begin{figure}[h!]
\centering
\includegraphics[width=0.8\columnwidth]{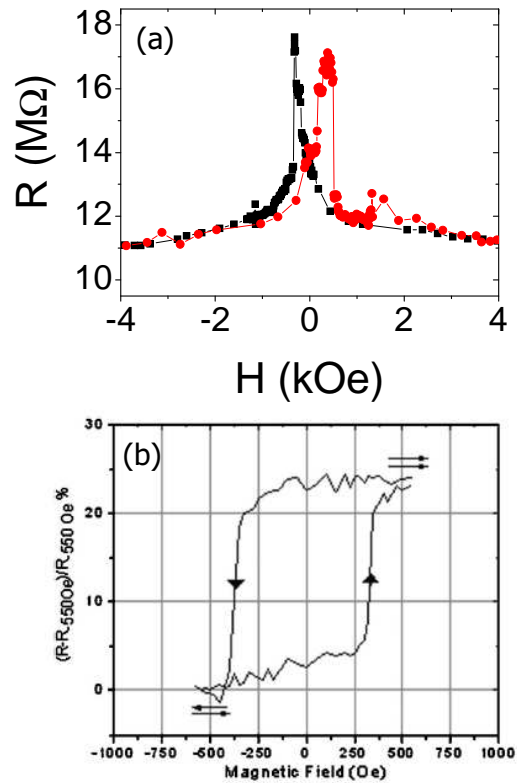}
\caption{(a) TMR curve measured at 4K and 10 mV on a Au/NiFe$_2$O$_4$/LSMO spin-filter (from \cite{luders2006}).
(b) Minor TMR loop measured at 300K on a Au/CoFe$_2$O$_4$/MgAl$_2$O$_4$/Fe$_3$O$_4$ spin-filter (from
\cite{chapline2006}).} \label{ferrites}
\end{figure}

\subsection{Spinel ferrites}

The spinel oxide family has been most widely studied. First experiments by the group of Suzuki using a
ferrimagnetic CoCr$_2$O$_4$ barrier in Fe$_3$O$_4$/CoCr$_2$O$_4$/La$_{0.7}$Sr$_{0.3}$MnO$_3$ tunnel junctions
did not clearly demonstrate a spin-filter effect by the barrier due to the use of two magnetic materials as
electrode and counter-electrode \cite{hu2002}. L\"uders \emph{et al} \cite{luders2006,luders2006a} observed a
large TMR effect of 52\% at 4K in Au/NiFe$_2$O$_4$/La$_{0.7}$Sr$_{0.3}$MnO$_3$ and
Au/NiFe$_2$O$_4$/SrTiO$_3$/La$_{0.7}$Sr$_{0.3}$MnO$_3$ tunnel junctions. In the latter heterostructures, the STO
spacer is introduced to induce a better magnetic decoupling between LSMO and NiFe$_2$O$_4$. Figure
\ref{ferrites}a represents the magnetoresistance curve of such a spin filter. When the configuration of the
magnetizations of the NiFe$_2$O$_4$ barrier and the LSMO counter-electrode goes from an antiparallel to a
parallel state, a large drop in the resistance is observed. The positive TMR observed, with a resistance larger
in the antiparallel state than in the parallel one corresponds to a positive spin filtering efficiency of 23\%.
This is in contrast with the negative spin polarization that could be expected from band structure calculations
(conduction band smaller for spin down than for spin up) for the inverse and the normal spinel structure
\cite{szotek2004,szotek2004b}. The positive spin filtering efficiency has been attributed to the different
symmetry of the spin up and spin down conduction band and to the resulting symmetry filtering \cite{luders2006}.
More recently, Chapline and Wang \cite{chapline2006} have claimed to have observed a sizeable filtering effect
at room temperature in Au/CoFe$_2$O$_4$/MgAl$_2$O$_4$/Fe$_3$O$_4$. The MgAl$_2$O$_4$ layer has been introduced
to decouple the magnetizations of the spin-filter barrier (CoFe$_2$O$_4$) and the electrode (Fe$_3$O$_4$).
Measurements are performed by using a conducting tip AFM. Fig. \ref{ferrites}b shows the minor loop
magnetoresistance curve obtained while reversing the Fe$_3$O$_4$ magnetization. A positive magnetoresistance of
24\% is observed at room temperature. The corresponding positive filtering efficiency has been attributed to the
lower energy of the t$_{2g}$ conduction band of the Co$^{2+}$ ions in antisites. Such a change of sign in the
filtering efficiency by cation inversion is in contrast with the band structure calculations of Szotek \emph{et
al} \cite{szotek2004,szotek2004b} that suggest the preservation of a minority gap smaller than the majority one
in the normal spinel structure as for the inverse spinel structure of NiFe$_2$O$_4$.

\subsection{BiMnO$_3$}

Large TMR effects have also been reported for barriers of some ferromagnetic-insulating manganites (BiMnO$_3$
\cite{gajek2005} and La$_{0.1}$Bi$_{0.9}$MnO$_3$ (LBMO) \cite{gajek2006a,gajek2006}). Due to the multiferroic
character of these materials results are reported in the following Part.

We end this Part on spin-filtering by mentioning that the bias-dependence of the TMR in spin-filters has not
been properly addressed yet. Few papers report on experimental results or on calculations
\cite{saffarzadeh2004}. A very rapid decrease of the TMR at low bias followed by a less pronounced variation
have been reported by L\"uders \emph{et al} \cite{luders2006} and Gajek \emph{et al} \cite{gajek2005}, which
could be due to excitation of spin waves in the ferromagnetic insulating barrier as found with NiO barriers
\cite{tsui71}. Theoretical input of this idea has not been published yet.

\section{Multiferroics}
\label{partV}

\begin{figure}
\centering
\includegraphics[width=\columnwidth]{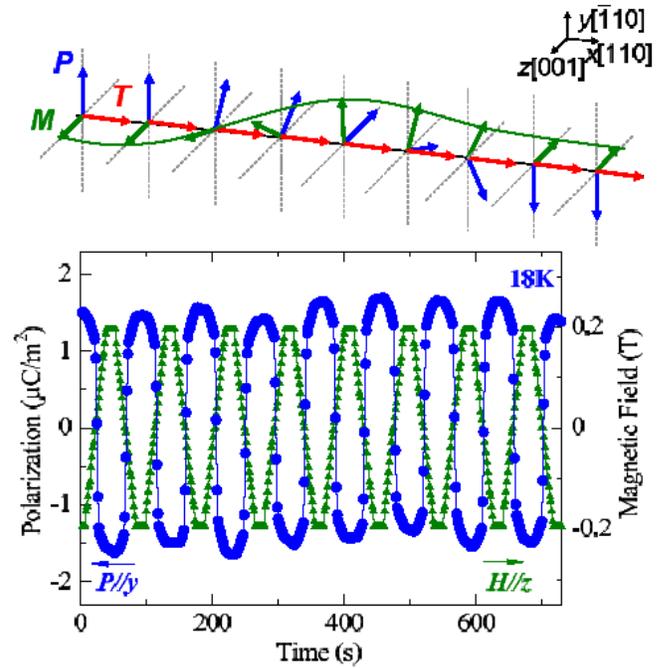}
\caption{Reproducible reversal of the ferroelectric polarization by an external magnetic field in the
spin-spiral compound CoCr$_2$O$_4$. The upper panel shows the geometrical relationship between M, P and T
(toroidal moment), from \cite{yamasaki2006}.} \label{CoCr2O4}
\end{figure}

Multiferroics is an emerging family of materials in the field of spintronics. These multifunctional materials
present two or more ferroic orders among ferromagnetic, ferroelectric, ferroelastic or ferrotoroidic.
Magnetoelectric-multiferroics in which ferromagnetic (but it is also a general trend to include antiferromagnets
or weak ferromagnets in this definition) and ferroelectric orders coexist with a magnetoelectric coupling
between them, open the possibility of controlling the polarization by a magnetic field or the magnetization by
an electric field. This should allow the design of, for example, ferroelectric memories which can be
magnetically recorded or, more interestingly, to MRAM recorded electrically. Whereas the reversal of the
polarization by a magnetic field or even the induction of a ferroelectric order by a magnetic field have been
reported in several materials (see for example \cite{kimura2005,kimura2005a,hur2004,yamasaki2006} and figure
\ref{CoCr2O4}), literature is extremely scarce as to the reciprocal effect, which may be due to the small number
of ferro- (or ferri-) magnetic ferroelectrics. The small abundance of single-phase multiferroic materials
\cite{hill2000} is circumvented by the emerging field of artificial multiferroics combining ferroelectric and
magnetic materials in a same heterostructure \cite{zhang2004,zavaliche2005}.

Most of the work reported on multiferroics has been performed on bulk materials but an increasing effort is made
to obtain high-quality thin films. The purpose of this chapter is not to make an exhaustive review on
multiferroics but to highlight a few experiments on thin films of interest in the field of spintronics. For
further information on multiferroic materials, the reader is referred to recent reviews
\cite{fiebig2005,prellier2005,eerenstein2006}.

The most studied multiferroic material is BiFeO$_3$, which is one of the few multiferroics with both critical
temperatures (i.e. magnetic and ferroelectric) much larger than 300K \cite{smolenskii63,fischer80}. Its
antiferromagnetic \cite{eerenstein2005,wang2005,bea2005} character has been used to induce an exchange bias on a
soft ferromagnetic layer at room temperature \cite{dho2006,bea2006a}. This capability, together with the
preservation of the ferroelectric character of this material down to 2 nm \cite{bea2006} as well as the
observation of a tunnel magnetoresistance through BiFeO$_3$ barriers \cite{bea2006a} make this material very
promising in the field of spintronics and should allow the design of tunnel magnetoresistance-based devices
controlled electrically \cite{binek2005}.

\begin{figure}
\centering
\includegraphics[width=\columnwidth]{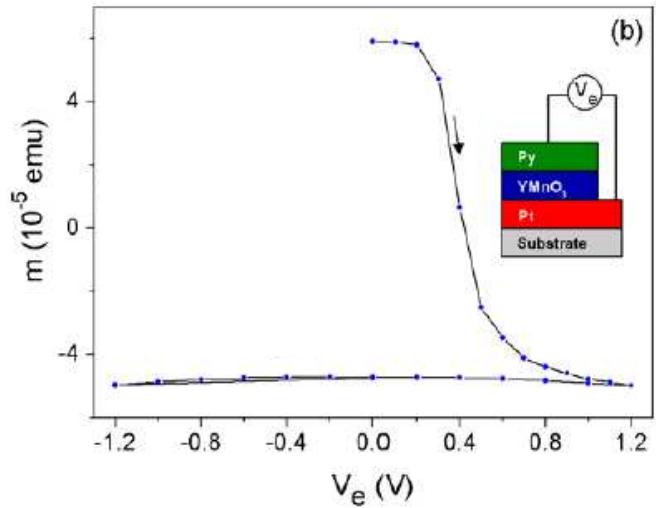}
\caption{Magnetization of a Ni$_{80}$Fe$_{20}$/YMnO$_3$ bilayer, in which an exchange bias is induced on
Ni$_{80}$Fe$_{20}$ by YMnO$_3$, as a function of the voltage applied across the YMnO$_3$ layer (from
\cite{laukhin2006}).} \label{YMnO3}
\end{figure}

The potential of multiferroic materials is further illustrated by two very recent results. The first one is the
first demonstration of the electric field control of the exchange bias in hexagonal YMnO$_3$ based
heterostructures \cite{laukhin2006}. The antiferromagnetic character of the compound has been used to exchange
bias a thin SrRuO$_3$ \cite{marti2006} or Ni$_{80}$Fe$_{20}$ \cite{laukhin2006}. Applying a dc voltage on the
antiferromagnetic-ferroelectric YMnO$_3$ layer causes a modification of the exchange bias and reversal of the
magnetization of the Ni$_{80}$Fe$_{20}$ layer in a Ni$_{80}$Fe$_{20}$/YMnO$_3$/Pt heterostructure, as shown on
Fig. \ref{YMnO3}.

The exploitation of the multifunctional character of these multiferroic materials is also exemplified by the
recent results of Gajek and coworkers in BiMnO$_3$ and LBMO-based heterostructures \cite{gajek2006}. BiMnO$_3$
is an established ferromagnetic insulator \cite{bokov66,sugawara68,atou99}. In addition, a ferroelectric
character has been claimed by several groups \cite{moreira2002b,son2004,sharan2004}. Gajek \emph{et al} have
proposed to use a ferromagnetic and ferroelectric material as a tunnel barrier in order to obtain a four
resistance level device. Thin ferromagnetic insulating layers of BiMnO$_3$ and La$_{0.1}$Bi$_{0.9}$MnO$_3$ have
been used as ferromagnetic tunnel barriers in Au/La$_{1-x}$Bi$_x$MnO$_3$/LSMO spin filters
\cite{gajek2005,gajek2006a} with 22\% and 35\% spin filtering efficiency respectively. By applying a magnetic
field, switching the configuration of the magnetizations of the barrier and the LSMO counter-electrode from
antiparallel to parallel produced two distinct resistance states related to the tunnel magnetoresistance effect.
The polarization of the ferroelectric La$_{0.1}$Bi$_{0.9}$MnO$_3$ barrier gives an additional degree of freedom
and accordingly two other resistance states are obtained, corresponding to an electroresistance phenomena of
20\% \cite{gajek2006} (for details on ferroelectric tunnel junctions, see \cite{tsymbal2006}).

\section{Spintronics devices with oxide materials}

Research on half-metallic oxides has allowed to measure record TMR values \cite{bowen2003} and to get a better
understanding of spin-dependent tunneling. However, manganite-based tunnel junctions show a vanishingly small
TMR at 300K and are thus not usable for applications. Room temperature results using magnetic oxides with higher
T$_C$ such as Fe$_3$O$_4$ are modest \cite{seneor99}, with TMR ratios one order of magnitude smaller than those
obtained on last-generation MgO-based MTJs \cite{lee2006b}. However, some very recent developments permit some
optimism as to the possible future use of magnetic oxides in spintronics devices.

A first important observation is that the vast majority of magnetic oxides with high critical temperatures are
(i) Fe-based and (ii) insulating. One might thus argue that future device-oriented research should focus on
these systems. Such insulating Fe oxides exist in many structural families (perovskites, garnets, spinels,
corundum, etc) and can be antiferromagnetic or ferrimagnetic. So far, mostly two groups of insulating Fe oxides
have been studied for spintronics purposes: spinel ferrites as spin-filter barriers and BiFeO$_3$ (and derived
compounds) as multiferroic elements.

Spin-filters have the potential to generate currents with spin-polarization approaching 100\%, which seems to be
difficult by using epitaxial diamagnetic barriers such as MgO, because of the finite tunneling probability for
states of all symmetries. If such very highly spin-polarized current sources become available, for example using
spinel ferrite tunnel barriers, the range of possible applications for spintronics devices might extend rapidly
to logic in addition to data storage.

Another promising family of materials for spintronics applications are multiferroics, and especially BiFeO$_3$
that has very high critical temperatures and has already been studied extensively in thin film form. Since
BiFeO$_3$ is not ferromagnetic but antiferromagnetic (in addition to being ferroelectric), its interest for
spintronics could be for exchange-biasing purposes. Robust and large exchange bias effects has been recently
demontrated at room temperature in BiFeO$_3$/CoFeB bilayers \cite{bea2006b}. The combination of exchange bias
with magnetoelectric coupling \cite{zhao2006} might allow to switch electrically the resistance of a spin-valve
or MTJ adjacent to a BFO layer in the near future. This achievement would represent a major breakthrough for
spintronics since it would allow to write a magnetic bit electrically, with very little power consumption.
Questions might then arise as to the integration of devices based on multiferroics into CMOS technology.
Good-quality epitaxial BFO films can be grown onto Si substrates \cite{wang2004b}. In this regard, it is also
important to note that perovskite oxides have already been integrated to CMOS technology for information storage
applications. Ferroelectric memories (FERAMs) \cite{scott2000} using ferroelectric Pb(Zr,Ti)O$_3$ (PZT) have
been fabricated by Fujitsu using a 0.5 $\mu$m CMOS process and integrated as 4kbit memory elements in each Sony
Playstation 2 (that have been sold to over 100 millions units worldwide). Panasonic and Samsung are now shipping
32 Mb and 64 Mb FERAM chips, respectively.

Significantly, Fujistu announced in July 2006 that their next-generation FERAMs will be based on BiFeO$_3$, a
lead-free ferroelectric \cite{fujitsu2006}. A multiferroic material will thus be soon used in a wide-scale
microelectronic product. CMOS industry is thus already exploiting the potential of complex oxides for
applications and wide-scale fabrication processes to define active perovskite oxide layers with lateral
dimensions in the 100 nm range are available.

On a longer timescale, the development of reliable high T$_C$ diluted magnetic oxides with large
spin-polarization at room-temperature and above, and electrically tunable magnetic and electronic properties
might play a role in future spintronics devices. ZnO is a particularly promising candidate (due to its long spin
lifetime \cite{ghosh2005}) and prospectives spintronics applications based on ZnO would benefit from the huge
effort that is currently being made to use ZnO in optoelectronics and to fabricate low-dimension ZnO structures
\cite{ozgur2005}.

\section{Conclusions and perspectives}

The use of magnetic oxides in spintronics architectures dates back from only 10 years ago. Since that time, this
field has been developing quickly. Record spin-polarization and TMR values have been reported in several
systems, confirming the predicted half-metallic character of manganites, CrO$_2$ and SFMO. New families of
magnetic oxides, namely diluted magnetic oxides and multiferroics, have emerged and started to reveal their
potential for spintronics. Even though this has been a leitmotif goal for many years, the observation of a large
spin-polarization at room temperature has not been possible until now, while in the meantime epitaxial MgO-based
junctions have reached TMR values as high a 400 \% \cite{yuasa2006} at 300K. However, the recent discovery of
novel high-T$_C$ spin-polarized magnetic oxides \cite{toyosaki2005,herranz2006} and the observation of large
spin-filtering effects with spinel ferrites \cite{luders2006,chapline2006} bring hopes of achieving large TMR
effects at room temperature and beyond, using magnetic oxide materials.

In addition, the relevance of oxides for spintronics does not restrict to generating highly spin-polarized
currents. Approaches to exploit their multifunctional character are promising and reveal new or poorly addressed
physical phenomena. Efforts towards an electrical control of magnetization are being pursued by several original
approaches exploiting for instance the influence of carriers on magnetization in Co-doped TiO$_2$
\cite{zhao2005}, magnetoelectric coupling \cite{laukhin2006} or strain effects \cite{thiele2005}.
Current-induced magnetization switching, due to spin-transfer, has been observed in manganite-based tunnel
junctions, with a poor reproducibility \cite{sun99a}. This approach certainly deserves further attention, from
both experimentalists and theorists.

We also note that bilayers consisting of a ferroelectric/piezoelectric oxide and a ferromagnet (either oxide or
metal) are currently being investigated \cite{thiele2005}. Electric modulations of the magnetic properties of
the ferromagnet by the ferroelectric or piezoelectric layer have been reported
\cite{thiele2006,eerenstein2006b}. While most experimental work has focused on epitaxial perovskite bilayers
such as LSMO/BaTiO$_3$ \cite{eerenstein2006b}, important effects have also been predicted for BaTiO$_3$/Fe
\cite{duan2006} and should thus be observable at room temperature and above.

Other promising approaches for oxide-based electronics and spintronics rely on engineering interfaces between
two oxides, to design two-dimensional phases with novel electronic properties. An interesting example in
provided by the observation of a metallic behavior at a LaTiO$_3$/SrTiO$_3$ interface, i.e. a system combining a
Mott insulator (LaTiO$_3$) and a band insulator (SrTiO$_3$) \cite{ohtomo2002,takizawa2006}. Theoretically
calculations on this type of interfaces predict a rich variety of novel properties
\cite{okamoto2004,okamoto2005,okamoto2006,kancharla2006,thulasi2006,lee2006}, like a ferromagnetic state
\cite{okamoto2004}. Interesting charge-transfer effects have also been observed at LaAlO$_3$/SrTiO$_3$
\cite{ohtomo2004} interfaces, but the reported transport properties may be dominated by the conductive
electron-doped SrTiO$_3$ substrate \cite{herranz2006c}. The large Hall mobilities observed in such
SrTiO$_3$-based systems \cite{tufte67,herranz2006b} as well as in ZnO films \cite{tsukazaki2006} are promising
for spin-transport experiments in field-effect transistor samples \cite{ueno2003,takahashi2006}.

Future research directions in oxide spintronics may also exploit the new physical properties of oxides in
further reduced dimensions, i.e. one- and zero-dimensional objects. It is interesting to note that nanowires or
nanotubes of several magnetic oxides have been synthesized (Fe$_3$O$_4$ \cite{sui2004,liao2006}, LSMO
\cite{curiale2005}, Co-doped TiO$_2$ \cite{wu2005}, etc), with in some case highly unexpected and promising
properties \cite{krusin2004}. To perform spin-dependent transport in gated structures using these nanotubes
\cite{sahoo2005} will challenge experimentalists and might unveil novel physical effects. Optimism is possible
in view of the recent spin-dependent transport measurements in planar devices combining LSMO spin-injecting and
spin-detecting electrodes with carbon nanotubes \cite{hueso2006,hueso2006a}.


%
%

\section*{Acknowledgment}
The authors would like to thank J.M.D. Coey, A. Fert, V. Garcia, G. Herranz, N.D. Mathur, J.S. Moodera, F.
Petroff, T.S. Santos and J.F. Scott for helpful comments on the manuscript.

\begin{biographynophoto}{Manuel Bibes}
was born in 1976 in Sainte-Foy-la-Grande, France. He received his Ph.D. degree in 2001 from the Universitat
Aut\`onoma de Barcelona, Spain and the Institut National des Sciences Appliqu\'ees, Toulouse, France after a
joint Franco-Spanish thesis on interfaces in manganites at the Institut de Ci\`encia de Materials de Barcelona,
under the supervision of J. Fontcuberta and J.-C. Ousset. After a two-year postdoc in A. Fert's group in Orsay,
France he became a CNRS research scientist at the Institut d'Electronique Fondamentale, Orsay, France in 2003.
His current research interests include spin-dependent tunneling, spin-filtering, multiferroics and diluted
magnetic oxides.

\end{biographynophoto}


\begin{biographynophoto}{Agn\`es Barth\'el\'emy}
received her Ph.D. in 1991 from the University of Paris-XI, Orsay, France, under the supervision of A. Fert. She
became an assistant professor at this university that same year, carrying out her research activity in the
Unit\'e Mixte de Physique CNRS-Thomson. During her Ph.D. and the following years, her main research subject was
giant magnetoresistance in metallic multilayers. In 1998, she began working on magnetic tunnel junctions based
on magnetic oxides such as manganites. She became a Professor of Physics in 2004 and is now leading the
Multifunctional Oxides group at the Unit\'e Mixte de Physique CNRS-Thales. Her research focuses on the physics
of such oxides and their integration in spintronics architectures.

\end{biographynophoto}


\begin{thebibliography}{100}
\providecommand{\url}[1]{#1} \csname url@rmstyle\endcsname \providecommand{\newblock}{\relax}
\providecommand{\bibinfo}[2]{#2} \providecommand\BIBentrySTDinterwordspacing{\spaceskip=0pt\relax}
\providecommand\BIBentryALTinterwordstretchfactor{4}
\providecommand\BIBentryALTinterwordspacing{\spaceskip=\fontdimen2\font plus
\BIBentryALTinterwordstretchfactor\fontdimen3\font minus
  \fontdimen4\font\relax}
\providecommand\BIBforeignlanguage[2]{{%
\expandafter\ifx\csname l@#1\endcsname\relax
\typeout{** WARNING: IEEEtran.bst: No hyphenation pattern has been}%
\typeout{** loaded for the language `#1'. Using the pattern for}%
\typeout{** the default language instead.}%
\else \language=\csname l@#1\endcsname \fi #2}}

\bibitem{zutic2004}
I.~\v{Z}uti\'{c}, J.~Fabian, and S.~{Das Sarma}, \emph{Rev. Mod. Phys.},
  vol.~76, p. 323, 2004.

\bibitem{baibich88}
{M.N. Baibich}, {J.-M. Broto}, A.~Fert, {F. Nguyen Van Dau}, F.~Petroff,
  P.~Eitenne, G.~Creuzet, A.~Friedrich, and J.~Chazelas, \emph{Phys. Rev.
  Lett.}, vol.~61, p. 2472, 1988.

\bibitem{bednorz86}
{J.G. Bednorz} and {K.A. Mueller}, \emph{Z. Phys. B}, vol.~64, p. 189, 1986.

\bibitem{lu96}
Y.~Lu, W.~Li, G.~Gong, G.~Xiao, A.~Gupta, P.~Lecoeur, J.~Sun, Y.~Wang, and
  V.~Dravid, \emph{Phys. Rev. B}, vol.~54, p. R8357, 1996.

\bibitem{ohno96}
H.~Ohno, A.~Shen, F.~Matsukura, A.~Oiwa, A.~Endo, S.~Katsumoto, and Y.~Iye,
  \emph{Appl. Phys. Lett.}, vol.~69, p. 363, 1996.

\bibitem{julliere75}
M.~Julli\`ere, \emph{Phys. Lett.}, vol. 54A, p. 225, 1975.

\bibitem{moodera95}
J.~Moodera, L.~Kinder, T.~Wong, and R.~Meservey, \emph{Phys. Rev. Lett.},
  vol.~74, p. 3273, 1995.

\bibitem{tsymbal2003}
{E.Y. Tsymbal}, {O.N. Mryasiv}, and {P.R. LeClair}, \emph{J. Phys.: Condens.
  Matter}, vol.~15, p. R109, 2003.

\bibitem{fert2006}
A.~Fert, A.~Barth\'el\'emy, and F.~Petroff, \emph{Nanomagnetism, ultrathin
  films, multilayers and Nanostructures}, ser. Contemporary concepts of
  Condensed Matter Science, {D. L. Mills} and {J. A. C. Bland}, Eds.\hskip 1em
  plus 0.5em minus 0.4em\relax Amsterdam, the Netherlands: Elsevier, 2006.

\bibitem{tedrow70}
{P.M. Tedrow}, R.~Meservey, and P.~Fulde, \emph{Phys. Rev. Lett.}, vol.~25, p.
  1270, 1970.

\bibitem{tedrow71}
{P.M. Tedrow} and R.~Meservey, \emph{Phys. Rev. Lett.}, vol.~26, p. 192, 1971.

\bibitem{deteresa99a}
{J.M. de Teresa}, A.~Barth\'el\'elmy, A.~Fert, {J.-P. Contour}, F.~Montaigne,
  and P.~Seneor, \emph{Science}, vol. 286, p. 507, 1999.

\bibitem{deteresa99}
{J.M. de Teresa}, A.~Barth\'el\'elmy, A.~Fert, {J.-P. Contour}, R.~Lyonnet,
  F.~Montaigne, P.~Seneor, and A.~Vaur\`es, \emph{Phys. Rev. Lett.}, vol.~82,
  p. 4288, 1999.

\bibitem{tsymbal2000}
{E.Y. Tsymbal}, {I.I. Oleinik}, and {D.G. Pettifor}, \emph{J. Appl. Phys.},
  vol.~87, p. 5230, 2000.

\bibitem{oleinik2000}
{I.I. Oleinik}, {E.Y. Tsymbal}, and {D.G. Pettifor}, \emph{Phys. Rev. B},
  vol.~62, p. 3952, 2000.

\bibitem{oleinik2002}
------, \emph{Phys. Rev. B}, vol.~65, p. 020401, 2002.

\bibitem{mavropoulos2000}
P.~Mavropoulos, N.~Papanikolaou, and {P.H. Dederichs}, \emph{Phys. Rev. Lett.},
  vol.~85, p. 1088, 2000.

\bibitem{maclaren99}
{J.M. MacLaren}, {X.-G. Zhang}, {W.H. Butler}, and X.~Wang, \emph{Phys. Rev.
  B}, vol.~63, p. 054416, 2001.

\bibitem{butler2001}
{W.H. Butler}, {X.-G. Zhang}, {T.C. Schulthess}, and {J.M. MacLaren},
  \emph{Phys. Rev. B}, vol.~63, p. 054416, 2001.

\bibitem{faure-vincent2003}
{J. Faure-Vincent}, C.~Tiusan, E.~Jouguelet, F.~Canet, M.~Sajieddine,
  C.~Bellouard, E.~Popova, M.~Hehn, F.~Montaigne, and A.~Schuhl, \emph{Appl.
  Phys. Lett.}, vol.~82, p. 4507, 2003.

\bibitem{yuasa2004}
S.~Yuasa, A.~Fukushima, T.~Nagahama, K.~Ando, and Y.~Suzuki, \emph{Jpn. J.
  Appl. Phys.}, vol.~43, p. L588, 2004.

\bibitem{yuasa2004a}
S.~Yuasa, T.~Nagahama, A.~Fukushima, Y.~Suzuki, and K.~Ando, \emph{Nat.
  Mater.}, vol.~3, p. 868, 2004.

\bibitem{parkin2004}
{S.S.P. Parkin}, {C. Kaiser}, {A. Panchula}, {P.M. Rice}, {B. Hughes}, {M.
  Samant}, and {S.-H. Yang}, \emph{Nat. Mater.}, vol.~3, p. 862, 2004.

\bibitem{yuasa2006}
S.~Yuasa, A.~Fukushima, H.~Kubota, Y.~Suzuku, and K.~Ando, \emph{Appl. Phys.
  Lett.}, vol.~89, p. 042505, 2006.

\bibitem{coey2002}
J.~Coey and M.~Venkatesan, \emph{J. Appl. Phys.}, vol.~91, p. 8345, 2002.

\bibitem{coey2004}
{J.M.D. Coey} and S.~Sanvito, \emph{J. Phys. D: Appl. Phys.}, vol.~37, p. 988,
  2004.

\bibitem{degroot83}
{R.A. de Groot}, {F.M. Mueller}, {P.G. van Engen}, and {K.H.J. Buschow},
  \emph{Phys. Rev. Lett.}, vol.~50, p. 2024, 1983.

\bibitem{yanase84}
A.~Yanase and K.~Siratori, \emph{J. Phys. Soc. Jpn.}, vol.~53, p. 312, 1984.

\bibitem{schwartz86}
K.~Schwartz, \emph{J. Phys. F: Met. Phys.}, vol.~16, p. L211, 1986.

\bibitem{pickett96}
W.~Pickett and D.~Singh, \emph{Phys. Rev. B}, vol.~53, p. 1146, 1996.

\bibitem{park98b}
{J.H. Park}, E.~Vescovo, {H.J. Kim}, C.~Kwon, R.~Ramesh, and T.~Venkatesan,
  \emph{Nature (London)}, vol. 392, p. 794, 1998.

\bibitem{sun96}
{J.Z. Sun}, {W.J. Gallagher}, {P.R. Ducombe}, {L. Krusin-Elbaum}, {R.A.
  Altman}, A.~Gupta, Y.~Lu, {G.Q. Gong}, and G.~Xiao, \emph{Appl. Phys. Lett.},
  vol.~69, p. 3266, 1996.

\bibitem{viret97}
M.~Viret, M.~Drouet, J.~Nassar, {J.-P. Contour}, C.~Fermon, and A.~Fert,
  \emph{Europhys. Lett.}, vol.~39, p. 545, 1997.

\bibitem{bowen2003}
M.~Bowen, M.~Bibes, A.~Barth\'el\'emy, J.-P. Contour, A.~Anane,
  Y.~Lema\^{\i}tre, and A.~Fert, \emph{Appl. Phys. Lett.}, vol.~82, p. 233,
  2003.

\bibitem{imada98}
M.~Imada, A.~Fujimori, and Y.~Tokura, \emph{Rev. Mod. Phys.}, vol.~70, p. 1039,
  1998.

\bibitem{coey99}
J.~Coey, M.~Viret, and S.~von Molnár, \emph{Adv. in Phys.}, vol.~48, p. 167,
  1999.

\bibitem{salamon2001}
{M.B. Salamon} and M.~Jaime, \emph{Rev. Mod. Phys.}, vol.~73, p. 583, 2001.

\bibitem{ziese2002}
M.~Ziese, R.~Hohne, N.~Hong, J.~Dienelt, K.~Zimmer, and P.~Esquinazi, \emph{J.
  Magn. Magn. Mater.}, vol. 242-245, p. 450, 2002.

\bibitem{haghiri2003}
{A.-M. Haghiri-Gosnet} and {J.-P. Renard}, \emph{J. Phys. D: Appl. Phys.}, vol.
  376, p. R127, 2003.

\bibitem{dorr2006}
K.~D\"orr, \emph{J. Phys. D: Appl. Phys.}, vol.~39, p. R125, 2006.

\bibitem{sun97}
{J.Z. Sun}, L.~Krusin-Elbaum, {P.R. Duncombe}, A.~Gupta, and {R.B. Laibowitz},
  \emph{Appl. Phys. Lett.}, vol.~70, p. 1769, 1997.

\bibitem{jo2000}
M.~Jo, N.~Mathur, N.~Todd, and M.~Blamire, \emph{Phys. Rev. B}, vol.~61, p.
  R14905, 2000.

\bibitem{bibes2006}
V.~Garcia, M.Bibes, and A.~Barth\'el\'emy, unpublished.

\bibitem{lyu99}
P.~Lyu, {D.Y. Xing}, and J.~Dong, \emph{Phys. Rev. B}, vol.~60, p. 4235, 1999.

\bibitem{obata99}
T.~Obata, T.~Manako, Y.~Shimakawa, and Y.~Kubo, \emph{Appl. Phys. Lett.},
  vol.~74, p. 290, 1999.

\bibitem{odonnell2000a}
{J. O'Donnell}, {A.E. Andrus}, S.~Oh, {E.V. Colla}, and J.~Eckstein,
  \emph{Appl. Phys. Lett.}, vol.~76, p. 218, 2000.

\bibitem{park98a}
{J.H. Park}, E.~Vescovo, {H.-J. Kim}, C.~Kwon, R.~Ramesh, and T.~Venkatesan,
  \emph{Phys. Rev. Lett.}, vol.~81, p. 1953, 1998.

\bibitem{sun99}
J.~Sun, D.~Abraham, R.~Rao, and C.~Eom, \emph{Appl. Phys. Lett.}, vol.~74, p.
  3017, 1999.

\bibitem{bibes2001e}
M.~Bibes, {Ll. Balcells}, S.~Valencia, J.~Fontcuberta, M.~Wojcik, E.~Jedryka,
  and S.~Nadolski, \emph{Phys. Rev. Lett.}, vol.~87, p. 067210, 2001.

\bibitem{jo99}
{M.-H. Jo}, {N.D. Mathur}, {J.E. Evetts}, {M.G. Blamire}, M.~Bibes, and
  J.~Fontcuberta, \emph{Appl. Phys. Lett.}, vol.~75, p. 3689, 1999.

\bibitem{izumi2001}
M.~Izumi, Y.~Ogimoto, Y.~Okimoto, T.~Manako, P.~Ahmet, K.~Nakajima, T.~Chikyow,
  M.~Kawasaki, and Y.~Tokura, \emph{Phys. Rev. B}, vol.~64, p. 064429, 2001.

\bibitem{millis98b}
{A.J. Millis}, T.~Darling, and A.~Migliori, \emph{J. Appl. Phys.}, vol.~83, p.
  1588, 1998.

\bibitem{yamada2004}
H.~Yamada, Y.~Ogawa, Y.~Ishii, H.~Sato, M.~Kawasaki, H.~Akoh, and Y.~Tokura,
  \emph{Science}, vol. 305, p. 646, 2004.

\bibitem{bibes2002}
M.~Bibes, S.~Valencia, {Ll. Balcells}, B.~Martinez, J.~Fontcuberta, M.~Wojcik,
  S.~Nadolski, and E.~Jedryka, \emph{Phys. Rev. B}, vol.~66, p. 134416, 2002.

\bibitem{garcia2004}
V.~Garcia, M.~Bibes, A.~Barth\'el\'emy, M.~Bowen, E.~Jacquet, {J.-P. Contour},
  and A.~Fert, \emph{Phys. Rev. B}, vol.~69, p. 052403, 2004.

\bibitem{kumigashira2006}
H.~Kumigashira, A.~Chikamatsu, R.~Hashimoto, M.~Oshima, T.~Ohnishi, M.~Lippmaa,
  H.~Wadati, A.~Fujimori, K.~Ono, M.~Kawasaki, and H.~Koinuma, \emph{Appl.
  Phys. Lett.}, vol.~88, p. 192504, 2006.

\bibitem{ishii2006}
Y.~Ishii, H.~Yamada, H.~Sato, H.~akoh, Y.~Ogawa, M.~Kawasaki, and Y.~Tokura,
  \emph{Appl. Phys. Lett.}, vol.~89, p. 042509, 2006.

\bibitem{sun98}
J.~Sun, \emph{Phil. Trans. R. Soc. London A}, vol. 356, p. 1693, 1998.

\bibitem{bowen2005}
M.~Bowen, A.~Barth\'el\'emy, M.~Bibes, E.~Jacquet, {J.-P. Contour}, A.~Fert,
  F.~Ciccacci, L.~D\`uo, and R.~Bertacco, \emph{Phys. Rev. Lett.}, vol.~95, p.
  137203, 2005.

\bibitem{zhang2001}
J.~Zhang \emph{et~al.}, \emph{Phys. Rev. B}, vol.~64, p. 184404, 2001.

\bibitem{gu2001}
{R.Y. Gu}, L.~Sheng, and {C.S. Ting}, \emph{Phys. Rev. B}, vol.~63, p.
  220406(R), 2001.

\bibitem{bibes2006a}
M.Bibes, M.~Bowen, and A.~Barth\'el\'emy, unpublished.

\bibitem{bratkovsky97}
{A.M. Bratkovsky}, \emph{Phys. Rev. B}, vol.~56, p. 2344, 1997.

\bibitem{bertacco2002}
R.~Bertacco, M.~Portalupi, M.~Marcon, L.~D\`uo, F.~Ciccacci, M.~Bowen, {J.-P.
  Contour}, and A.~Barth\'el\'emy, \emph{J. Magn. Magn. Mater.}, vol. 242-245,
  p. 710, 2002.

\bibitem{sakuraba2006}
Y.~Sakuraba, M.~Hattori, M.~Oogane, Y.~Ando, H.~Kato, A.~Sakuma, T.~Mizayaki,
  and H.~Kubota, \emph{Appl. Phys. Lett.}, vol.~88, p. 192508, 2006.

\bibitem{bowen2007}
M.~Bowen \emph{et~al.}, to appear in J. Phys.: Condens. Matter.

\bibitem{hayakawa2002}
J.~Hayakawa, S.~Kokado, K.~Ito, M.~Sugiyama, H.Asano, M.~Matsui, A.~Sakuma, and
  M.~Ichimura, \emph{Jpn. J. Appl. Phys.}, vol.~41, p. 1340, 2002.

\bibitem{hayakawa2002a}
J.~Hayakawa, K.~Ito, S.~Kokado, M.~Ichimura, A.~Sakuma, M.~Sugiyama, H.Asano,
  and M.~Matsui, \emph{J. Appl. Phys.}, vol.~91, p. 8792, 2002.

\bibitem{bowen2006}
M.~Bowen, V.~Cros, H.~Jaffr\`es, P.~Bencok, F.~Petroff, and {N.B. Brookes},
  \emph{Phys. Rev. B}, vol.~73, p. 012405, 2006.

\bibitem{bibes2003a}
M.~Bibes, M.~Bowen, A.~Barth\'el\'emy, A.~Anane, K.~Bouzehouane,
  C.~Carr\'et\'ero, E.~Jacquet, {J.-P. Contour}, and O.~Durand, \emph{Appl.
  Phys. Lett.}, vol.~82, p. 3269, 2003.

\bibitem{garcia2005}
V.~Garcia, M.~Bibes, {J.-L. Maurice}, E.~Jacquet, K.~Bouzehouane, {J.-P.
  Contour}, and A.~Barth\'el\'emy, \emph{Appl. Phys. Lett.}, vol.~87, p.
  212501, 2005.

\bibitem{mathon2001}
J.~Mathon and A.~Umerski, \emph{Phys. Rev. B}, vol.~63, p. 220403(R), 2001.

\bibitem{zhang2004}
{X.-G. Zhang} and {W.H. Butler}, \emph{Phys. Rev. B}, vol.~70, p. 172407, 2004.

\bibitem{belashchenko2005}
{K.D. Belashchenko}, J.~Velev, and {E.Y. Tsymbal}, \emph{Phys. Rev. B},
  vol.~72, p. 140404(R), 2005.

\bibitem{velev2005}
{J.P. Velev}, {K.D. Belaschenko}, {D.A. Stewart}, {M. van Schilfgaarde}, {S.S.
  Jaswal}, and {E.Y. Tsymbal}, \emph{Phys. Rev. Lett.}, vol.~95, p. 216601,
  2005.

\bibitem{bowen2006a}
M.~Bibes, A.~Barth\'el\'emy, V.~Bellini, M.~Bibes, P.~Seneor, E.~Jacquet,
  {J.-P. Contour}, and {P.H. Dederichs}, \emph{Phys. Rev. B}, vol.~73, p.
  140408(R), 2006.

\bibitem{velev2006}
{J.P. Velev} \emph{et~al.}, private communication.

\bibitem{balcells2000}
{Ll. Balcells}, J.~Cifre, A.~Calleja, J.~Fontcuberta, M.~Varela, and
  F.~Benitez, \emph{Sensors and Actuators}, vol.~81, p.~64, 2000.

\bibitem{goyal97}
A.~Goyal, M.~Rajeswari, R.~Shreekala, {S.E. Lofland}, {S.M. Bhagat},
  T.~Boettcher, C.~Kwon, R.~Ramesh, and T.~Venkatesan, \emph{Appl. Phys.
  Lett.}, vol.~71, p. 2535, 1997.

\bibitem{yang2006}
{C.-H. Yang}, J.~Koo, C.~Song, {T.Y. Koo}, {K.-B. Lee}, and {Y.H. Jeong},
  \emph{Phys. Rev. B}, vol.~73, p. 224112, 2006.

\bibitem{bona85}
{G.L. Bona}, F.~Meier, M.~Taborelli, E.~Bucher, and {P.H. Schmidt}, \emph{Solid
  State Commun.}, vol.~56, p. 391, 1985.

\bibitem{patterson63}
F.~Patterson, C.~Moeller, and R.~Ward, \emph{Inorg. Chem.}, vol.~2, p. 196,
  1963.

\bibitem{galasso66}
F.~Galasso, {F.C. Douglas}, and {R.J. Kasper}, \emph{J. Chem. Phys.}, vol.~44,
  p. 1672, 1966.

\bibitem{nakagawa68}
T.~Nakagawa, \emph{J. Phys. Soc. Jpn.}, vol.~24, p. 806, 1968.

\bibitem{sleight62}
{A.W. Sleight}, J.~Longo, and R.~Ward, \emph{Inorg. Chem.}, vol.~1, p. 245,
  1962.

\bibitem{kobayashi98}
K.~Kobayashi, T.~Kimura, H.~Sawada, K.~Terakura, and Y.~Tokura, \emph{Nature},
  vol. 395, p. 677, 1998.

\bibitem{hwang96}
{H.Y. Hwang}, {S.-W. Cheong}, {N.P. Ong}, and B.~Batlogg, \emph{Phys. Rev.
  Lett.}, vol.~77, p. 2041, 1996.

\bibitem{sarma2000a}
{D.D. Sarma}, P.~Mahadevan, {T. Saha-Dasgupta}, S.~Ray, and A.~Kumar,
  \emph{Phys. Rev. Lett.}, vol.~85, p. 2549, 2000.

\bibitem{besse2002}
M.~Besse, V.~Cros, A.~Barth\'el\'emy, H.~Jaffr\`es, J.~Vogel, F.~Petroff,
  A.~Mirone, A.~Tagliaferri, P.~Bencok, P.~Decorse, P.~Berthet, Z.~Szotek,
  {W.M. Temmerman}, {S.S. Dhesi}, {N.B. Brookes}, A.~Rogalev, and A.~Fert,
  \emph{Europhys. Lett.}, vol.~60, p. 608, 2002.

\bibitem{navarro2001}
J.~Navarro, C.~Frontera, L.~Balcells, B.~Martinez, and J.~Fontcuberta,
  \emph{Phys. Rev. B}, vol.~64, p. 092411, 2001.

\bibitem{navarro2003}
J.~Navarro, J.~Nogu\`es, {J.S. Mu\~noz}, and J.~Fontcuberta, \emph{Phys. Rev.
  B}, vol.~67, p. 174416, 2003.

\bibitem{granado2002}
E.~Granado, Q.~Huang, {J.W. Lynn}, J.~Gopalakrishnan, {R.L. Greene}, and
  K.~Ramesha, \emph{Phys. Rev. B}, vol.~66, p. 064409, 2002.

\bibitem{kato2004}
H.~Kato, T.~Okuda, Y.~Okimoto, Y.~Tomioka, K.~Oikawa, T.~Kamiyama, and
  Y.~Yokura, \emph{Phys. Rev. B}, vol.~69, p. 184412, 2004.

\bibitem{kato2002}
H.~Kato, T.~Okuda, Y.~Okimoto, Y.~Yomioka, Y.~Takenoya, A.~Ohkubo, M.~Kawasaki,
  and Y.~Tokura, \emph{Appl. Phys. Lett.}, vol.~81, p. 328, 2002.

\bibitem{vaitheeswaran2005}
G.~Vaitheeswaran, V.~Kanchana, and A.~Delin, \emph{Appl. Phys. Lett.}, vol.~86,
  p. 032513, 2005.

\bibitem{tang2005}
{C.Q. Tang}, Y.~Zhang, and J.~Dai, \emph{Solid State Commun.}, vol. 133, p.
  219, 2005.

\bibitem{balcells2001}
{Ll. Balcells}, J.~Navarro, M.~Bibes, A.~Roig, B.~Martinez, and J.~Fontcuberta,
  \emph{Appl. Phys. Lett.}, vol.~78, p. 781, 2001.

\bibitem{navarro2001a}
J.~Navarro, {Ll. Balcells}, F.~Sandiumenge, M.~Bibes, A.~Roig, B.~Martinez, and
  J.~Fontcuberta, \emph{J. Phys.: Condens. Matter}, vol.~13, p. 8481, 2001.

\bibitem{sarma2000}
{D.D. Sarma}, {E.V. Sampathkumarana}, S.~Ray, R.~Nagarajan, S.~Majumdar,
  A.~Kumar, G.~Nalini, and {T.N. Guru Row}, \emph{Solid. State. Commun.}, vol.
  114, p. 465, 2000.

\bibitem{navarro2003a}
J.~Navarro, C.~Frontera, D.~Rubi, N.~Mestres, and J.~Fontcuberta, \emph{Mater.
  Res. Bull.}, vol.~38, p. 1477, 2003.

\bibitem{rager2002}
J.~Rager, {A.V. Berenov}, {L.F. Cohen}, {W.R. Branford}, {Y.V. Bugoslavsky},
  {Y. Miyoshi}, {M. Ardakani}, and {J.L. MacManus-Driscoll}, \emph{Appl. Phys.
  Lett.}, vol.~81, p. 5003, 2002.

\bibitem{asano2001}
H.~Asano, M.~Osugi, Y.~Kohara, D.~Higashida, and M.~Matsui, \emph{Jpn. J. Appl.
  Phys.}, vol.~40, p. 4883, 2001.

\bibitem{manako99}
T.~Manako, M.~Izumi, Y.~Konishi, {K.I. Kobayashi}, M.~Kawasaki, and Y.~Tokura,
  \emph{Appl. Phys. Lett.}, vol.~74, p. 2215, 1999.

\bibitem{venimadhav2004}
A.~Venimadhav, F.~Sher, {J.P. Attfield}, and {M.G. Blamire}, \emph{J. Magn.
  Magn. Mater.}, vol. 269, p. 101, 2004.

\bibitem{sanchez2004}
D.~S\'anchez, M.~Garcia-Hern\'andez, N.~Auth, and G.~Jakob, \emph{J. Appl.
  Phys.}, vol.~96, p. 2736, 2004.

\bibitem{fix2005}
T.~Fix, G.~Versini, {J.L. Loison}, S.~Colis, G.~Schmerber, G.~Pourroy, and
  A.~Dinia, \emph{J. Appl. Phys.}, vol.~97, p. 024907, 2005.

\bibitem{wang2006}
S.~Wang, H.~Pan, X.~Zhang, G.~Lian, and G.~Xion, \emph{Appl. Phys. Lett.},
  vol.~88, p. 121912, 2006.

\bibitem{santiso2002}
J.~Santiso, A.~Figueras, and J.~Fraxedas, \emph{Surf. Interface Anal.},
  vol.~33, p. 676, 2002.

\bibitem{besse2002a}
M.~Besse, F.~Pailloux, A.~Barth\'el\'emy, K.~Bouzehouane, A.~Fert, J.~Olivier,
  O.~Durand, F.~Wyczisk, R.~Bisaro, and {J.-P. Contour}, \emph{J. Cryst.
  Growth}, vol. 241, p. 448, 2002.

\bibitem{bibes2003}
M.~Bibes, K.~Bouzehouane, M.~Besse, A.~Barth\'el\'emy, S.Fusil, M.~Bowen,
  P.~Seneor, J.-P. Contour, and A.~Fert, \emph{Appl. Phys. Lett.}, vol.~83, p.
  2629, 2003.

\bibitem{asano2005}
H.~Asano, N.~Koduka, K.~Imaeda, M.~Sugiyama, and M.~Matsui, \emph{IEEE Trans.
  Magn.}, vol.~41, p. 2811, 2005.

\bibitem{auth2003}
N.~Auth, G.~Jakob, T.~Block, and C.~Felser, \emph{Phys. Rev. B}, vol.~68, p.
  024403, 2003.

\bibitem{bugoslavsky2005}
Y.~Bugoslavsky, Y.~Miyoshi, {S.K. Clowes}, {W.R. Branford}, M.~Lake, I.~Brown,
  {A.D. Caplin}, and {L.F. Cohen}, \emph{Phys. Rev. B}, vol.~71, p. 104523,
  2005.

\bibitem{bouzehouane2003}
K.~Bouzehouane \emph{et~al.}, \emph{Nanoletters}, vol.~3, p. 1599, 2003.

\bibitem{philipp2003}
{J.B. Philipp}, D.~Reisinger, M.~Schonecke, M.~Opel, A.~Marx, A.~Erb, L.~Alff,
  and R.~Gross, \emph{J. Appl. Phys.}, vol.~93, p. 6853, 2003.

\bibitem{venimadhav2006}
A.~Venimadhav, F.~Sher, {J.P. Attfield}, and {M.G. Blamire}, \emph{Solid State
  Commun.}, vol. 138, p. 314, 2006.

\bibitem{asano2004}
H.~Asano, N.~Nozuka, A.~Tsuzuki, and M.~Matsui, \emph{Appl. Phys. Lett.},
  vol.~85, p. 263, 2004.

\bibitem{serrate2007}
D.~Serrate, {J.M. de Teresa}, and {M.R. Ibarra}, \emph{J. Phys.: Condens.
  Matter}, vol.~9, p. 023201, 2007.

\bibitem{watts2000}
{S.M. Watts}, S.~Wirth, {S. von Moln\`ar}, A.~Barry, and {J.M.D. Coey},
  \emph{Phys. Rev. B}, vol.~61, p. 9621, 2000.

\bibitem{li99}
{X.W. Li}, A.~Gupta, {T.R. McGuire}, {P.R. Duncombe}, and G.~Xiao, \emph{J.
  Appl. Phys.}, vol.~85, p. 5585, 1999.

\bibitem{ivanov2001}
{P.G. Ivanov}, {S.M. Watts}, and {D.M. Lind}, \emph{J. Appl. Phys.}, vol.~89,
  p. 1035, 2001.

\bibitem{ranno97}
L.~Ranno, A.~Barry, and {J.M.D. Coey}, \emph{J. Appl. Phys.}, vol.~81, p. 5774,
  1997.

\bibitem{shima2002}
M.~Shima, T.~Tepper, and {C.A Ross}, \emph{J. Appl. Phys.}, vol.~91, p. 7920,
  2002.

\bibitem{coey98a}
{J.M.D. Coey}, {A.E. Berkowitz}, {Ll. Balcells}, {F.F. Putris}, and A.~Barry,
  \emph{Phys. Rev. Lett.}, vol.~80, p. 3815, 1998.

\bibitem{soulen98}
{R.J. Soulen}, {J.M. Byers}, {M.S. Osofsky}, B.~Nadgorny, T.~Ambrose, {S.F.
  Cheng}, {P.R. Broussard}, {C.T. Tanaka}, J.~Nowak, {J.S. Moodera}, A.~Barry,
  and {J.M.D. Coey}, \emph{Science}, vol. 282, p.~85, 1998.

\bibitem{desisto2000}
{W.J. DeSisto}, {P.R. Broussard}, {B.E. Nadgorny}, and {M.S. Osofsky},
  \emph{Appl. Phys. Lett.}, vol.~76, p. 3789, 2000.

\bibitem{anguelouch2001}
{A. Anguelouch}, A.~Gupta, G.~Xiao, {D.W. Abraham}, Y.~Ji, S.~Ingvarsson, and
  {C.L. Chien}, \emph{Phys. Rev. B}, vol.~64, p. 180408(R), 2001.

\bibitem{parker2002}
{J.S. Parker}, {S.M. Watts}, {P.G. Ivanov}, and P.~Xiong, \emph{Phys. Rev.
  Lett.}, vol.~88, p. 196601, 2002.

\bibitem{barry2000}
A.~Barry, {J.M.D. Coey}, and M.~Viret, \emph{J. Phys.: Condens. Matter},
  vol.~12, p. L173, 2000.

\bibitem{gupta2001}
A.~Gupta, {X.W. Li}, and G.~Xiao, \emph{Appl. Phys. Lett.}, vol.~78, p. 1894,
  2001.

\bibitem{parker2004}
{J.S. Parker}, {P.G. Ivanov}, {D.M. Lind}, P.~Xiong, and {Y. Xin}, \emph{Phys.
  Rev. B}, vol.~69, p. 220413(R), 2004.

\bibitem{astrov60}
{D.N. Astrov}, \emph{Sov. Phys. JETP}, vol.~11, p. 708, 1960.

\bibitem{loos2002}
J.~Loos and P.~Nov\'ak, \emph{Phys. Rev. B}, vol.~66, p. 132403, 2002.

\bibitem{degroot86}
{R.A. de Groot} and {K.H.J. Buschow}, \emph{J. Magn. Magn. Mater.}, vol. 54-57,
  p. 1377, 1986.

\bibitem{zhang91}
Z.~Zhang and S.~Satpathy, \emph{Phys. Rev. B}, vol.~44, p. 13319, 1991.

\bibitem{antonov2003}
{V.N. Antonov}, {B.N. Harmon}, and {A.N. Yaresko}, \emph{Phys. Rev. B},
  vol.~67, p. 024417, 2003.

\bibitem{seneor99}
P.~Seneor, A.~Fert, J.-L. Maurice, F.~Montaigne, F.~Petroff, and A.~Vaur\`es,
  \emph{Appl. Phys. Lett.}, vol.~74, p. 4017, 1999.

\bibitem{alldredge2006}
{L.M.B. Alldredge}, {R.V. Chopdekar}, {B.B. Nelson-Cheeseman}, and Y.~Suzuki,
  \emph{J. Appl. Phys.}, vol.~99, p. 08K303, 2006.

\bibitem{todo95}
S.~Todo, K.~Siratori, and S.~Kimura, \emph{J. Phys. Soc. Jpn.}, vol.~64, p.
  2118, 1995.

\bibitem{walz2002}
F.~Walz, \emph{J. Phys.: Condens. Matter}, vol.~14, p. R285, 2002.

\bibitem{garcia2004a}
J.~Garcia and G.~Subias, \emph{J. Phys.: Condens. Matter}, vol.~16, p. R145,
  2004.

\bibitem{margulies96}
{D.T. Margulies}, {F.T. Parker}, {F.E. Spada}, {R.S. Goldman}, J.~Li,
  R.~Sinclair, and {A.E. Berkowitz}, \emph{Phys. Rev. B}, vol.~53, p. 9175,
  1996.

\bibitem{voogt95}
{F.C. Voogt}, T.~Hibma, {G.L. Zhang}, M.~Hoefman, and L.~Niesen, \emph{Surf.
  Sci.}, vol. 331-333, p. 1508, 1995.

\bibitem{gong97}
{G.Q. Gong}, A.~Gupta, G.~Xiao, W.~Qian, and {V.P. Dravid}, \emph{Phys. Rev.
  B}, vol.~56, p. 5096, 1997.

\bibitem{margulies97}
{D.T. Margulies}, {F.T. Parker}, {M.L. Rudee}, {F.E. Spada}, {J.N. Chapman},
  {P.R. Aitchison}, and {A.E. Berkowitz}, \emph{Phys. Rev. Lett.}, vol.~79, p.
  5162, 1997.

\bibitem{voogt98}
{F.C. Voogt}, {T.T.M. Palstra}, L.~Niesen, {O.C. Rogojanu}, {M.A. James}, and
  T.~Hibma, \emph{Phys. Rev. B}, vol.~57, p. R8107, 1998.

\bibitem{eerenstein2003}
W.~Eerenstein, L.~Kalev, L.~Nielsen, {T.T.M. Palstra}, and T.~Hibma, \emph{J.
  Magn. Magn. Mater.}, vol. 258-259, p.~73, 2003.

\bibitem{li98}
{X.W. Li}, A.~Gupta, G.~Xiao, and G.~Gong, \emph{J. Appl. Phys.}, vol.~83, p.
  7049, 1998.

\bibitem{eerenstein2002}
W.~Eerenstein, {T.T.M. Palstra}, T.~Hibma, and S.~Celotto, \emph{Phys. Rev. B},
  vol.~66, p. 201101, 2002.

\bibitem{eerenstein2003a}
W.~Eerenstein, ``Spin-dependent transport accross anti-phase boundaries in
  magnetite thin films,'' Ph.D. dissertation, Rijksuniveristeit Groningen, the
  Netherlands, 2003.

\bibitem{huang2002}
{D.J. Huang}, {C.F. Chang}, J.~Chen, {L.H. Tjeng}, {A.D. Rata}, {W.P. Wu},
  {S.C. Chung}, {H.J. Lin}, T.~Hibma, and {C.T. Chen}, \emph{J. Magn. Magn.
  Mater.}, vol. 239, p. 261, 2002.

\bibitem{huang2002a}
{D.J. Huang}, {L.H. Tjeng}, J.~Chen, {C.F. Chang}, {W.P. Wu}, {A.D. Rata},
  T.~Hibma, {S.C. Chung}, {S.-G. Chyu}, {C.-C. Wu}, and {C.T. Chen},
  \emph{Surf. Rev. Lett.}, vol.~9, p. 1007, 2002.

\bibitem{dedkov2002}
{Yu.S. Dedkov}, U.~R\"udiger, and G.~G\"untherodt, \emph{Phys. Rev. B},
  vol.~65, p. 064417, 2002.

\bibitem{fonin2003}
M.~Fonin, {Y.S. Dedkov}, J.~Mayer, U.~R\"udiger, and G.~G\"untherodt,
  \emph{Phys. Rev. B}, vol.~68, p. 045414, 2003.

\bibitem{bataille2006}
{A.M. Bataille}, A.~Tagliaferri, S.~Gota, {C. de Nada\"i}, {J.-B. Moussy},
  {M.-J. Guittet}, K.~Bouzehouane, F.~Petroff, {M. Gautier-Soyer}, and {N.B.
  Brookes}, \emph{Phys. Rev. B}, vol.~73, p. 172201, 2006.

\bibitem{fonin2005}
M.~Fonin, R.~Pentcheva, {Y.S. Dedkov}, M.~Sperlich, {D.V. Vyalikh},
  M.~Scheffler, U.~R\"udiger, and G.G\"untherodt, \emph{Phys. Rev. B}, vol.~72,
  p. 104436, 2005.

\bibitem{aoshima2003}
K.~Aoshima and {S.X. Wang}, \emph{J. Appl. Phys.}, vol.~93, p. 7954, 2003.

\bibitem{yoon2004}
{K.S. Yoon}, {J.H. Koo}, {Y.H. Do}, {K.W. Kim}, {C.O. Kim}, and {J.P. Hong},
  \emph{J. Magn. Magn. Mater.}, vol. 285, p. 125, 2004.

\bibitem{bataille2006a}
{A.M. Bataille}, R.~Mattana, P.~Seneor, A.~Tagliaferri, S.~Gota,
  K.~Bouzehouane, C.~Deranlot, {M.-J. Guillet}, {J.-B. Moussy}, {C. de
  Nada\"i}, {N.B. Brookes}, F.~Petroff, and {M. Gautier-Soyer}, to be published
  in J. Magn. Magn. Mater.

\bibitem{reisinger2004}
D.~Reisinger, P.~Majewski, M.~Opel, L.~Alff, and R.~Gross, condmat/0407725.

\bibitem{hu2002}
G.~Hu and Y.~Suzuki, \emph{Phys. Rev. Lett.}, vol.~89, p. 276601, 2002.

\bibitem{alldredge2006b}
{L.M.B. Alldredge}, {R.V. Chopdekar}, {B.B. Nelson-Cheeseman}, and Y.~Suzuki,
  \emph{Appl. Phys. Lett.}, vol.~89, p. 182504, 2006.

\bibitem{parkin2006}
{S.S.P. Parkin}, unpublished.

\bibitem{yoon2004a}
{K.S. Yoon}, {J.Y. Yang}, {K.W. Kim}, {J.H. Koo}, {C.O. Kim}, and {J.P. Hong},
  \emph{J. Appl. Phys.}, vol.~95, p. 6933, 2004.

\bibitem{vandijken2004}
{S. van Dijken}, X.~Fain, {S.M. Watts}, K.~Nakajima, and {J.M.D. Coey},
  \emph{J. Magn. Magn. Mater.}, vol. 280, p. 322, 2004.

\bibitem{snoeck2006}
E.~Snoeck, C.~Gatel, R.~Serra, G.~BenAssayag, {J.-B. Moussy}, {A.M. Bataille},
  M.~Pannetier, and {M. Gautier-Soyer}, \emph{Phys. Rev. B}, vol.~73, p.
  104434, 2006.

\bibitem{kennedy99}
{R.J. Kennedy} and {P.A. Stampe}, \emph{J. Phys. D: Appl. Phys.}, vol.~32,
  p.~16, 1999.

\bibitem{lu2004}
{Y.X. Lu}, {J.S. Claydon}, {Y.B. Xu}, {D.M. Schofield}, and {S.M. Thompson},
  \emph{J. Appl. Phys.}, vol.~95, p. 7228, 2004.

\bibitem{watts2004}
{S.M. Watts}, K.~Nakajima, {S. van Dijken}, and {J.M.D. Coey}, \emph{J. Appl.
  Phys.}, vol.~95, p. 7465, 2004.

\bibitem{reisinger2003}
D.~Reisinger, M.~Schonecke, T.~Brenninger, M.~Opel, A.~Erb, L.~Alff, and
  R.~Gross, \emph{J. Appl. Phys.}, vol.~94, p. 1854, 2003.

\bibitem{berdunov2004}
N.~Berdunov, S.~Murphy, G.~Mariotto, and {I.V. Shvets}, \emph{Phys. Rev.
  Lett.}, vol.~93, p. 057201, 2004.

\bibitem{miyamoto94}
Y.~Miyamoto, \emph{Ferroelectrics}, vol. 161, p. 117, 1994.

\bibitem{matsubara2005}
M.~Matsubara, Y.~Shimada, T.~Arima, Y.~Taguchi, and Y.~Tokua, \emph{Phys. Rev.
  B}, vol.~72, p. 220404(R), 2005.

\bibitem{austin70}
{I.G. Austin} and {D. Elwell}, \emph{Contempt. Phys.}, vol.~11, p. 455, 1970.

\bibitem{brabers95}
{V.A.M. Brabers}, \emph{Handbook of Magnetic Materials}, E.~Science, Ed., 1995,
  vol.~8.

\bibitem{szotek2004}
Z.~Szotek, \emph{Presented at the International Conference on Nanospintronics,
  Kyoto, May 24-28}, 2004.

\bibitem{zuo2006}
X.~Zuo, S.~Yan, B.~Barbiellini, {V.G. Harris}, and C.~Vittoria, \emph{J. Magn.
  Magn. Mater.}, vol. 303, p. e432, 2006.

\bibitem{itoh2006}
H.~Itoh, private communication.

\bibitem{luders2005}
U.~L\"uders, M.~Bibes, {J.-F. Bobo}, M.~Cantoni, R.~Bertacco, and
  J.~Fontcuberta, \emph{Phys. Rev. B}, vol.~71, p. 134419, 2005.

\bibitem{luders2006a}
U.~L\"uders, A.~Barth\'el\'emy, M.~Bibes, K.~Bouzehouane, S.~Fusil, E.~Jacquet,
  {J.-P. Contour}, {J.-F. Bobo}, {J. Fontcuberta}, and {A. Fert}, \emph{Adv.
  Mater.}, vol.~18, p. 1733, 2006.

\bibitem{luders2006c}
U.~L\"uders \emph{et~al.}, unpublished.

\bibitem{mccurrie94}
\emph{Ferromagnetic materials: structure and properties}.\hskip 1em plus 0.5em
  minus 0.4em\relax Academic Press, 1995.

\bibitem{venzke96}
S.~Venzke, R.~van Dover, J.~Philips, E.~Gyorgy, T.~Siegrist, C.-H. Chen,
  D.~Werder, R.~Fleming, R.~Felder, E.~Coleman, and R.~Opila, \emph{J. Mater.
  Res}, vol.~11, p. 1187, 1996.

\bibitem{yang2005}
A.~Yang, Z.~Chen, X.~Zuo, D.~Arena, J.~Kirkland, C.~Vittoria, and {V.G.
  Harris}, \emph{Appl. Phys. Lett.}, vol.~86, p. 252210, 2005.

\bibitem{kim2002}
{K.J. Kim}, {H.S. Lee}, {M.H. Lee}, and {S.H. Lee}, \emph{J. Appl. Phys.},
  vol.~91, p. 9974, 2002.

\bibitem{chinnasamy2001}
{C.N. Chinnasamy}, {A. Narayanasamy}, N.~Ponpandian, K.~Chattopadhyay,
  K.~Shinoda, B.~Jeyadevan, K.~Tohji, K.~Nakatsuka, T.~Furubayashi, and
  I.~Nakatani, \emph{Phys. Rev. B}, vol.~63, p. 184108, 2001.

\bibitem{zhou2002}
{Z.H. Zhou}, {J.M. Xue}, J.~Wang, {H.S.O. Chan}, T.~Yu, and {Z.X. Shen},
  \emph{J. Appl. Phys.}, vol.~91, p. 6015, 2002.

\bibitem{luders2006b}
U.~L\"uders, G.~Herranz, M.~Bibes, K.~Bouzehouane, E.~Jacquet, {J.-P. Contour},
  S.~Fusil, {J.-F. Bobo}, {J. Fontcuberta}, A.~Barth\'el\'emy, and {A. Fert},
  \emph{J. Appl. Phys.}, vol.~99, p. 08K301, 2006.

\bibitem{ishikawa2005}
M.~Ishikawa, H.~Tanaka, and T.~Kawai, \emph{Appl. Phys. Lett.}, vol.~86, p.
  222504, 2005.

\bibitem{tanaka2006}
H.~Tanaka \emph{et~al.}, unpublished.

\bibitem{eom92}
{C.B. Eom}, {R.J. Cava}, {R.M. Fleming}, {J.M. Phillips}, {R.B. van Dover},
  {J.H. Marshall}, {J.W.P. Hsu}, {J.J. Krajewski}, and {W.F. Peck Jr.},
  \emph{Science}, vol. 258, p. 1766, 1992.

\bibitem{allen96}
{P.B. Allen}, H.~Berger, O.~Chauvet, L.~Forro, T.~Jarlborg, A.~Junod, B.~Revaz,
  and G.~Santi, \emph{Phys. Rev. B}, vol.~53, p. 4393, 1996.

\bibitem{mazin97}
I.~Mazin and {D.J. Singh}, \emph{Phys. Rev. B}, vol.~56, p. 2556, 1997.

\bibitem{kiyama99}
T.~Kiyaman, K.~Yoshimura, K.~Kosuge, H.~Mitamara, and T.~Goto, \emph{J. Phys.
  Soc. Jpn.}, vol.~68, p. 3372, 1999.

\bibitem{singh96}
{D.J. Singh}, \emph{J. Appl. Phys.}, vol.~79, p. 4818, 1996.

\bibitem{worledge2000}
D.~Worledge and T.~Geballe, \emph{Appl. Phys. Lett.}, vol.~76, p. 900, 2000.

\bibitem{nadgorny2003}
B.~Nadgorny, {M.S. Osofsky}, {D.J. Singh}, {G.T. Woods}, {R.J. Soulen Jr.},
  {M.K. Lee}, {S.D. Bu}, and {C.B. Eom}, \emph{Appl. Phys. Lett.}, vol.~82, p.
  427, 2003.

\bibitem{takahashi2003}
{K.S. Takahashi}, A.~Sawa, Y.~Ishii, H.~Akoh, M.~Kawasaki, and Y.~Tokura,
  \emph{Phys. Rev. Lett.}, vol.~67, p. 094413, 2003.

\bibitem{noh2004}
{J.S. Noh}, {C.B. Eom}, {M.G. Lagally}, {J.Z. Sun}, and {H.C. Kim}, \emph{Phys.
  Stat. Sol. (b)}, vol. 241, p. 1490, 2004.

\bibitem{bibes99b}
M.~Bibes, B.~Mart\'{\i}nez, J.~Fontcuberta, V.~Trt\'{\i}k, F.~Ben\'{\i}tez,
  C.~Ferrater, F.~Sanch\'ez, and M.~Varela, \emph{Phys. Rev. B}, vol.~60, p.~1,
  1999.

\bibitem{dietl2000}
T.~Dietl, H.~Ohno, F.~Matsukura, J.~Cibert, and D.~Ferrand, \emph{Science},
  vol. 287, p. 1019, 2000.

\bibitem{jungwirth2006}
T.~Jungwirth, J.~Sinova, J.~Masek, J.~Kucera, and {A.H. MacDonald},
  condmat/0603380.

\bibitem{prellier2003}
W.~Prellier, A.~Fouchet, and B.~Mercey, \emph{J. Phys.: Condens. Matter},
  vol.~15, p. R1583, 2003.

\bibitem{pearton2004}
{S.J. Pearton}, {W.H. Heo}, M.~Ivill, {D.P. Norton}, and T.~Steiner,
  \emph{Semicond. Sci. Technol.}, vol.~19, p. R59, 2004.

\bibitem{janisch2005}
R.~Janisch, P.~Gopal, and {N.A. Spaldin}, \emph{J. Phys.: Condens. Matter},
  vol.~17, p. R657, 2005.

\bibitem{chambers2006}
S.~Chambers, \emph{Surf. Sci. Rep.}, vol.~61, p. 345, 2006.

\bibitem{matsumoto2001}
Y.~Matsumoto, M.~Murakami, T.~Shono, T.~Hasegawa, T.~Fukumura, M.~Kawasaki,
  P.~Ahmet, T.~Chikyow, S.~Koshihara, and H.~Koinuma, \emph{Science}, vol. 291,
  p. 854, 2001.

\bibitem{ueda2001}
K.~Ueda, H.~Tabata, and T.~Kawai, \emph{Appl. Phys. Lett.}, vol.~79, p. 988,
  2001.

\bibitem{saeki2001}
H.~Saeki, H.~Tabata, and T.~Kawai, \emph{Solid State Commun.}, vol. 120, p.
  439, 2001.

\bibitem{coey2005a}
{J.M.D. Coey}, M.~Venkatesan, and {C.B. Fitzgerald}, \emph{Nat. Mater.},
  vol.~4, p. 173, 2005.

\bibitem{coey2005}
{J.M.D. Coey}, \emph{Sol. State Sci.}, vol.~7, p. 660, 2005.

\bibitem{chambers2001}
{S.A. Chambers}, S.~Thevuthasan, {R.F.C. Farrow}, {R.F. Mark}, {J.U. Thiele},
  L.~Folks, {M.G. Samant}, {A.J. Kellock}, N.~Ruzycki, {D.L. Ederer}, and
  U.~Diebold, \emph{Appl. Phys. Lett.}, vol.~79, p. 3467, 2001.

\bibitem{chambers2002}
{S.A. Chambers}, {C.M. Wang}, S.~Thevuthasan, T.~Droubay, {D.E. McCready},
  {A.S. Lea}, V.~Shutthanandan, and {C.F. Windisch Jr.}, \emph{Thin Solid
  Films}, vol. 418, p. 197, 2002.

\bibitem{park2002}
{W.K. Park}, {R.J. Ortega-Hertogs}, {J.S. Moodera}, A.~Punnoose, and {M.S.
  Seehra}, \emph{J. Appl. Phys.}, vol.~91, p. 8093, 2002.

\bibitem{matsumoto2001a}
Y.~Matsumoto, R.~Takahashi, M.~Murakami, T.~Koida, {X.-J. Fan}, T.~Hasegawa,
  T.~Fukumura, M.~Kawasaki, {S.-Y. Koshihara}, and H.~Koinuma, \emph{Jpn. J.
  Appl. Phys.}, vol.~40, p. L1204, 2001.

\bibitem{kim2003}
{J.-Y. Kim}, {J.-H. Park}, {B.-G. Park}, {H.-J. Noh}, {S.-J. Oh}, {J.S. Yang},
  {D.-H. Kim}, {S.D. Bu}, {T.-W. Noh}, {H.-J. Lin}, {H.-H. Hsieh}, and {C.T.
  Chen}, \emph{Phys. Rev. Lett.}, vol.~90, p. 017401, 2003.

\bibitem{chambers2003}
{S.A. Chambers}, T.~Droubay, {C.M. Wang}, {A.S. Lea}, {R.F.C. Farrow},
  L.~Folks, V.~Deline, and S.~Anders, \emph{Appl. Phys. Lett.}, vol.~82, p.
  1257, 2003.

\bibitem{shinde2004}
{S.R. Shinde}, {S.B. Ogale}, {J.S Higgins}, H.~Zheng, {A.J. Millis}, {V.N.
  Kulkarni}, R.~Ramesh, {R.L. Greene}, and T.~Venkatesan, \emph{Phys. Rev.
  Lett.}, vol.~92, p. 166601, 2004.

\bibitem{griffin2006}
{K.A. Griffin}, M.~Varela, {S.J. Pennycook}, {A.B. Pakhomov}, and K.~Krishnan.

\bibitem{park2002a}
{M.S. Park}, {S.K. Kwon}, and {B.I. Min}, \emph{Phys. Rev. B}, vol.~65, p.
  161201(R), 2002.

\bibitem{janisch2006}
R.~Janisch and {N.A. Spaldin}, \emph{Phys. Rev. B}, vol.~73, p. 035201, 2006.

\bibitem{shinde2003}
{S.R. Shinde}, {S.B. Ogale}, {S. Das Sarma}, {J.R. Simpson}, {H.D. Drew}, {S.E.
  Lofland}, C.~Lanci, {J.B. Buban}, {N.D. Browning}, {V.N. Kulkarni},
  J.~Higgins, {R.P. Sharma}, {R.L. Greene}, and T.~Venkatesan, \emph{Phys. Rev.
  B}, vol.~67, p. 115211, 2003.

\bibitem{mamiya2006}
K.~Mamiya, T.~Koide, A.~Fujimori, H.~Tokano, H.~Manaka, A.~Tanaka, H.~Toyosaki,
  F.~Fukumura, and M.~Kawasaki, condmat/06003149.

\bibitem{toyosaki2004}
H.~Toyosaki, T.~Fukumura, Y.~Yamada, K.~Nakajima, T.~Chikyow, T.~Hasegawa,
  H.~Koinuma, and M.~Kawasaki, \emph{Nat. Mater.}, vol.~3, p. 221, 2004.

\bibitem{quilty2006}
{J.W. Quilty}, A.~Shibata, {J.-Y. Son}, K.~Takubo, T.~Mizokawa, H.~Toyosaki,
  T.~Fukumura, and M.~Kawasaki, \emph{Phys. Rev. Lett.}, vol.~96, p. 027202,
  2006.

\bibitem{jaffe2005}
{J.E. Jaffe}, {T.C. Droubay}, and {S.A. Chambers}, \emph{J. Appl. Phys.},
  vol.~97, p. 073908, 2005.

\bibitem{kim2002a}
{D.H. Kim}, {J.S. Yang}, {K.W. Lee}, {S.D. Bu}, {T.W. Noh}, {S.-J. Oh}, {Y.-W.
  Kim}, {J.-S. Chung}, H.~Tanaka, {H.Y. Lee}, and T.~Kawai, \emph{Appl. Phys.
  Lett.}, vol.~81, p. 2421, 2002.

\bibitem{yoon2006}
{S.D. Yoon}, Y.~Chen, A.~Yang, {T.L. Goodrich}, X.~Zuo, {D.A. Arena},
  K.~Zeimer, C.~Vittoria, and {V.G. Harris}, \emph{J. Phys.: Condens. Matter},
  vol.~18, p. L355, 2006.

\bibitem{kaspar2005}
{T.C. Kaspar}, {S.M. Heald}, {C.M. Wang}, {J.D. Bryan}, T.~Droubay,
  V.~Shutthanandan, S.~Thevuthasan, {D.E. McCready}, {A.J. Kellock}, {D.R.
  Gamelin}, and {S.A. Chambers}, \emph{Phys. Rev. Lett.}, vol.~97, p. 217203,
  2005.

\bibitem{toyosaki2005}
H.~Toyosaki, T.~Fukumura, K.~Ueno, M.~Nakano, and M.~Kawasaki, \emph{Jpn. J.
  Appl. Phys.}, vol.~44, p. L896, 2005.

\bibitem{toyosaki2006}
------, \emph{J. Appl. Phys.}, vol.~99, p. 08M102, 2006.

\bibitem{zhao2005}
T.~Zhao, {S.R. Shinde}, {S.B. Ogale}, H.~Zheng, Venkatesan, R.~Ramesh, and {S.
  Das Sarma}, \emph{Phys. Rev. Lett.}, vol.~94, p. 126601, 2005.

\bibitem{ozgur2005}
U.~\"Ozg\"ur, {Y.I. Alivov}, C.~Liu, A.~Teke, {M.A. Reshchikov}, S.~Dogan,
  V.~Avrutin, {S.-J. Cho}, and H.~Morko\c, \emph{J. Appl. Phys.}, vol.~98, p.
  041301, 2005.

\bibitem{kittilstved2006a}
{K.R. Kittilstved}, {W.K. Liu}, and {D.R. Gamelin}, \emph{Nat. Mater.}, vol.~5,
  p. 291, 2006.

\bibitem{maurice2006}
{J.-L. Maurice}, K.~Rode, A.~Anane, D.~Imhoff, and {J.-P. Contour}, \emph{Eur.
  Phys. J. B}, vol.~33, p. 109, 2006.

\bibitem{kundaliya2004}
{D.C. Kundaliya}, {S.B. Ogale}, {S.E. Lofland}, S.~Dhar, {C.J. Metting}, {S.F.
  Shinde}, Z.~Ma, B.~Varughese, {K.V. Ramanujachary}, {L. Salamanca-Riba}, and
  T.~Venkatesan, \emph{Nat. Mater.}, vol.~3, p. 709, 2004.

\bibitem{ando2001}
K.~Ando, H.~Saito, Z.~Jin, T.~Fukumura, M.~Kawasaki, Y.~Matsumoto, and
  H.~Koinuma, \emph{J. Appl. Phys.}, vol.~89, p. 7284, 2001.

\bibitem{kittilstved2005}
{K.R. Kittilstved}, {N.S. Norberg}, and {D.R. Gamelin}, \emph{Phys. Rev.
  Lett.}, vol.~94, p. 147209, 2005.

\bibitem{kittilstved2006}
{K.R. Kittilstved}, {D.A. Schwartz}, {A.C. Tuan}, {S.M. Heald}, {S.A.
  Chambers}, and {D.R. Gamelin}, \emph{Phys. Rev. Lett.}, vol.~97, p. 137203,
  2006.

\bibitem{wang2004}
Q.~Wang, Q.~Sun, P.~Jena, and Y.~Kawazoe, \emph{Phys. Rev. B}, vol.~70, p.
  052408, 2004.

\bibitem{spaldin2004}
{N.A. Spaldin}, \emph{Phys. Rev. B}, vol.~69, p. 125201, 2004.

\bibitem{xu2006}
Q.~Xu, L.~Hartmann, H.~Schmidt, H.~Hochmuth, M.~lorenz, {R. Schmidt-Grund},
  C.~Sturm, D.~Spemann, and M.~Grundmann, \emph{Phys. Rev. B}, vol.~73, p.
  205342, 2006.

\bibitem{peng2006}
{Y.Z. Peng}, T.~Liew, {T.C. Chong}, {C.W. An}, and {D.W. Song}, \emph{Appl.
  Phys. Lett.}, vol.~88, p. 192110, 2006.

\bibitem{rode2006}
K.~Rode, ``Contribution \`a l\'etude des semiconducteurs magn\'etiques: cas des
  films minces d'oxyde de zinc dop\'e au cobalt,'' Ph.D. dissertation,
  Universit\'e Paris-Sud, Orsay, France, 2006.

\bibitem{fujimori92}
A.~Fujimori, I.~Hase, M.~Nakamura, Y.~Fujishima, Y.~Tokura, M.~Abbate, {F.M.F
  de Groot}, {M.T. Czyzyk}, {J.C. Fuggle}, O.~Strebel, F.~Lopez, M.~Domke, and
  G.~Kaindl, \emph{Phys. Rev. B}, vol.~46, p. 9841, 1992.

\bibitem{tokura93}
Y.~Tokura, Y.~Taguchi, Y.~Okada, Y.~Fujishima, T.~Arima, K.~Kumagai, and
  Y.~Iye, \emph{Phys. Rev. Lett.}, vol.~70, p. 2126, 1993.

\bibitem{sunstrom92}
{J.E. Sunstrom IV}, {S.M. Kauzlarich}, and P.~Klavins, \emph{Chem. Mater.},
  vol.~4, p. 346, 1992.

\bibitem{furukawa99}
Y.~Furukawa, I.~Okamura, K.~Kumagai, T.~Goto, T.~Fukase, Y.~Taguchi, and
  Y.~Tokura, \emph{Phys. Rev. B}, vol.~59, p. 10550, 1999.

\bibitem{wu2000}
W.~Wu, F.~Lu, {K.H. Wong}, G.~Pang, {C.L. Choy}, and Y.~Zhang, \emph{J. Appl.
  Phys.}, vol.~88, p. 700, 2000.

\bibitem{cho2001}
{J.H. Cho} and {H.J. Cho}, \emph{J. Appl. Phys.}, vol.~79, p. 1426, 2001.

\bibitem{zhao2003}
{Y.G. Zhao}, {S.R. Shinde}, {S.B. Ogale}, J.~Higgins, {R.J. Choudhary}, {V.N.
  Kulkarni}, {R.L. Greene}, T.~Venkatesan, {S.E. Loflanf}, C.~Lanci, {J.P.
  Buban}, {N.D. Browning}, {S. Das Sarma}, and {A.J. Millis}, \emph{Appl. Phys.
  Lett.}, vol.~83, p. 2199, 2003.

\bibitem{qiao2004}
{P.T. Qiao}, {Z.H. Zhao}, {Y.G. Zhao}, {X.P. Zhang}, {W.Y. Zhang}, {S.B.
  Ogale}, {S.R. Shinde}, T.~Venkatesan, {S.E. Lofland}, and C.~Lanci,
  \emph{Thin Solid Films}, vol. 468, p.~8, 2004.

\bibitem{ranchal2005}
R.~Ranchal, M.~Bibes, A.~Barth\'el\'emy, S.~Guyard, E.~Jacquet, {J.-P.
  Contour}, C.~Pascanut, P.~Berthet, and N.~Dragoe, \emph{J. Appl. Phys.},
  vol.~98, p. 013514, 2005.

\bibitem{herranz2006}
G.~Herranz, R.~Ranchal, M.~Bibes, H.~Jaffr\`es, E.~Jacquet, {J.-L. Maurice},
  K.~Bouzehouane, F.~Wyczisk, E.~Tafra, M.~Basletic, A.~Hamzic, C.~Colliex,
  {J.-P. Contour}, A.~Barth\'el\'emy, and A.~Fert, \emph{Phys. Rev. Lett.},
  vol.~96, p. 027207, 2006.

\bibitem{petroff2006}
F.~Petroff \emph{et~al.}, unpublished.

\bibitem{herranz2006a}
G.~Herranz \emph{et~al.}, unpublished.

\bibitem{zhang2006}
{S.X. Zhang}, {S.B. Ogale}, {D.C. Kundaliya}, {L.F. Fu}, {N.D. Browning},
  S.~Dhar, W.~Ramadan, {J.S. Higgins}, {R.L. Greene}, and T.~Venkatesan,
  \emph{Appl. Phys. Lett.}, vol.~89, p. 012501, 2006.

\bibitem{inaba2005}
J.~Inaba and T.~Katsufuji, \emph{Phys. Rev. B}, vol.~72, p. 052408, 2005.

\bibitem{iwasawa2006}
H.~Iwasawa, K.~Yamakawa, T.~Saitoh, J.~Inaba, T.~Katsufuji, M.~Higashigushi,
  K.~Shimada, H.~Namatame, and M.~Taniguchi, \emph{Phys. Rev. Lett.}, vol.~96,
  p. 067203, 2006.

\bibitem{esaki67}
L.~Esaki, {P.J. Stiles}, and S.~von Moln\`ar, \emph{Phys. Rev. Lett.}, vol.~19,
  p. 852, 1967.

\bibitem{moodera88}
{J.S. Moodera}, X.~Hao, {G.A. Gibson}, and R.~Meservey, \emph{Phys. Rev.
  Lett.}, vol.~61, p. 637, 1988.

\bibitem{leclair2002}
P.~LeClair, {J.K. Ha}, {H.J.M. Swagten}, {J.T. Kohlhepp}, {C.H. van de Vin},
  and {W.J.M. de Jonge}, \emph{Appl. Phys. Lett}, vol.~80, p. 625, 2002.

\bibitem{santos2004}
{T.S. Santos} and {J.S. Moodera}, \emph{Phys. Rev. B}, vol.~69, p. 241203(R),
  2004.

\bibitem{luders2006}
U.~L\"uders, M.~Bibes, K.~Bouzehouane, E.~Jacquet, {J.-P. Contour}, S.~Fusil,
  {J.-F. Bobo}, {J. Fontcuberta}, A.~Barth\'el\'emy, and {A. Fert}, \emph{Appl.
  Phys. Lett.}, vol.~88, p. 082505, 2006.

\bibitem{chapline2006}
{M.G. Chapline} and {S.X. Wang}, \emph{Phys. Rev. B}, vol.~74, p. 014418, 2006.

\bibitem{gajek2005}
M.~Gajek, M.~Bibes, A.~Barth\'el\'emy, K.~Bouzehouane, S.~Fusil, M.~Varela,
  J.~Fontcuberta, and A.~Fert, \emph{Phys. Rev. B}, vol.~72, p. 020406(R),
  2005.

\bibitem{gajek2006a}
{M. Gajek}, M.~Bibes, M.~Varela, J.~Fontcuberta, G.~Herranz, S.~Fusil,
  K.~Bouzehouane, A.~Barth\'el\'emy, and A.~Fert, \emph{J. Appl. Phys.},
  vol.~99, p. 08E504, 2006.

\bibitem{gajek2006}
M.~Gajek, M.~Bibes, S.~Fusil, K.~Bouzehouane, J.~Fontcuberta,
  A.~Barth\'el\'emy, and A.~Fert, condmat/0606444.

\bibitem{szotek2004b}
Z.~Szotek, {W.M. Temmerman}, A.~Svane, L.~Petit, P.~Strange, {G.M. Stocks},
  D.~K\"odderitzsch, W.~Hergert, and H.~Winter, \emph{J. Phys.: Condens.
  Matter}, vol.~16, p. S5587, 2004.

\bibitem{saffarzadeh2004}
A.~Saffarzadeh, \emph{J. Magn. Magn. Mater.}, vol. 269, p. 327, 2004.

\bibitem{tsui71}
{D.C. Tsui}, {R.E. Dietz}, and {L.R. Walker}, \emph{Phys. Rev. Lett.}, vol.~27,
  p. 1729, 1971.

\bibitem{kimura2005}
T.~Kimura, G.~Lawes, T.~Goto, Y.~Tokura, and {A.P. Ramirez}, \emph{Phys. Rev.
  B}, vol.~71, p. 224425, 2005.

\bibitem{kimura2005a}
T.~Kimura, G.~Lawes, and {A.P. Ramirez}, \emph{Phys. Rev. Lett.}, vol.~94, p.
  137201, 2005.

\bibitem{hur2004}
N.~Hur, S.~Park, {P.A. Sharma}, {J.S. Ahn}, S.~Guha, and {S.-W. Cheong},
  \emph{Nature}, vol. 429, p. 393, 2004.

\bibitem{yamasaki2006}
Y.~Yamasaki, S.~Miasaka, Y.~Kanko, {J.-P. He}, T.~Arima, and Y.~Tokura,
  \emph{Phys. Rev. Lett.}, vol.~96, p. 207204, 2006.

\bibitem{hill2000}
{N.A. Hill}, \emph{J. Phys. Chem. B}, vol. 104, p. 6694, 2000.

\bibitem{zavaliche2005}
F.~Zavaliche, H.~Zheng, {L. Mohaddes-Ardabili}, {S.Y. Yang}, Q.~Zhan,
  P.~Shafer, E.~Reilly, R.~Chopdekar, Y.~Jia, P.~Wright, {D.G. Schlom},
  Y.~Suzuki, and R.~Ramesh, \emph{Nanolett.}, vol.~5, p. 1793, 2005.

\bibitem{fiebig2005}
M.~Fiebig, \emph{J. Phys. D: Appl. Phys.}, vol.~38, p. R123, 2005.

\bibitem{prellier2005}
W.~Prellier, {M.P. Singh}, and P.~Murugavel, \emph{J. Phys.: Condens. Matter},
  vol.~17, p. R803, 2005.

\bibitem{eerenstein2006}
W.~Eerenstein, {N.D. Mathur}, and {J.F. Scott}, \emph{Nature}, vol. 442, p.
  759, 2006.

\bibitem{smolenskii63}
{G.A Smolenskii}, {V.M. Yudin}, {E.S. Sher}, and {Y.E. Stolypin}, \emph{Sov.
  Phys. JETP}, vol.~16, p. 622, 1963.

\bibitem{fischer80}
P.~Fischer, M.~Polomska, I.~Sosnowska, and M.~Szymanski, \emph{J. Phys. C},
  vol.~13, p. 1931, 1980.

\bibitem{eerenstein2005}
W.~Eerenstein, {F.D. Morrison}, J.~Dho, {M.G. Blamire}, and {J.F. Scott},
  \emph{Science}, vol. 307, p. 1203a, 2005.

\bibitem{wang2005}
J.~Wang, A.~Scholl, H.~Zheng, {S.B. Ogale}, D.~Viehland, {D.G. Schlom}, {K.M.
  Rabe}, M.~Wuttig, L.~Mohaddes, J.~Neaton, U.~Waghmare, T.~Zhao, and
  R.~Ramesh, \emph{Science}, vol. 307, p. 1203b, 2005.

\bibitem{bea2005}
H.~B{\'e}a, M.~Bibes, A.~Barth\'el\'emy, K.~Bouzehouane, E.~Jacquet, A.~Khodan,
  {J.-P. Contour}, S.~Fusil, F.~Wyczisk, A.~Forget, D.~Lebeugle, D.~Colson, and
  M.~Viret, \emph{Appl. Phys. Lett.}, vol.~87, p. 072508, 2005.

\bibitem{dho2006}
J.~Dho, X.~Qi, H.~Kim, {J.L. MacManus-Driscoll}, and {M.G. Blamire}, \emph{Adv.
  Mater.}, vol.~18, p. 1445, 2006.

\bibitem{bea2006a}
H.~B\'ea, M.~Bibes, S.~Cherifi, F.~Nolting, B.~Warot-Fonrose, S.~Fusil,
  G.~Herranz, C.~Deranlot, E.~Jacquet, K.~Bouzehouane, and A.~Barth\'el\'emy,
  \emph{Appl. Phys. Lett.}, vol.~89, p. 242114, 2006.

\bibitem{bea2006}
H.~B{\'e}a, S.~Fusil, K.~Bouzehouane, M.~Bibes, M.~Sirena, G.~Herranz,
  E.~Jacquet, {J.-P. Contour}, and A.~Barth\'el\'elmy, \emph{Jpn. J. Appl.
  Phys.}, vol.~45, p. L187, 2006.

\bibitem{binek2005}
C.~Binek and B.~Doudin, \emph{J. Phys.: Condens. Matter}, vol.~17, p. L39,
  2005.

\bibitem{laukhin2006}
V.~Laukhin, V.~Skumryev, X.~Marti, D.~Hrabovsky, F.~Sanchez, {M.V.
  Garcia-Cuenca}, C.~Ferrater, M.~Varela, U.~L\"uders, {J.-F. Bobo}, and
  J.~Fontcuberta, \emph{Phys. Rev. Lett.}, vol.~97, p. 227201, 2006.

\bibitem{marti2006}
X.~Marti, F.~Sanchez, J.~Fontcuberta, {M.V. Garcia-Cuenca}, C.~Ferrater, and
  M.~Varela, \emph{J. Appl. Phys.}, vol.~99, p. 08P302, 2006.

\bibitem{bokov66}
{V.A. Bokov}, {I.E. Myl'nikova}, {S.A. Kizhaev}, {M.F. Bryzhina}, and {N.A.
  Grigoryan}, \emph{Sov. Phys. Sol. State}, vol.~7, p. 2993, 1966.

\bibitem{sugawara68}
F.~Sugawara, S.~Iida, Y.~Syono, and A.~Akimoto, \emph{J. Phys. Soc. Jpn.},
  vol.~25, p. 1553, 1968.

\bibitem{atou99}
T.~Atou, H.~Chiba, K.~Ohoyama, Y.~Yamaguchi, and Y.~Syono, \emph{J. Solid State
  Chem}, vol. 145, p. 639, 1999.

\bibitem{moreira2002b}
{A. Moreira dos Santos}, S.~Parashar, {A.R. Raju}, {Y.S. Zhao}, {A.K.
  Cheetham}, and {C.N.R. Rao}, \emph{Solid State Commun.}, vol. 122, p.~49,
  2002.

\bibitem{son2004}
{J.Y. Son}, {B.G. Kim}, {C.H. Kim}, and {J.H. Cho}, \emph{Appl. Phys. Lett.},
  vol.~84, p. 4971, 2004.

\bibitem{sharan2004}
A.~Sharan, J.~Lettieri, Y.~Jia, W.~Tian, X.~Pan, {D.G. Schlom}, and V.~Gopalan,
  \emph{Phys. Rev. B}, vol.~69, p. 214109, 2004.

\bibitem{tsymbal2006}
{E.Y. Tsymbal} and H.~Kohlstedt, \emph{Science}, vol. 313, p. 181, 2006.

\bibitem{lee2006b}
{Y.M. Lee}, {J. Hayakawa}, S.~Ikeda, F.~Matsukura, and H.~Ohno, \emph{Appl.
  Phys. Lett.}, vol.~89, p. 04256, 2006.

\bibitem{bea2006b}
H.~B\'ea, M.~Bibes, S.~Fusil, K.~Bouzehouane, E.~Jacquet, K.~Rode, P.~Bencok,
  and A.~Barth\'el\'emy, \emph{Phys. Rev. B}, vol.~74, p. 020101(R), 2006.

\bibitem{zhao2006}
T.~Zhao, A.~Scholl, F.~Zavaliche, K.~Lee, M.~Barry, A.~Doran, {M.P. Cruz},
  {Y.H. Chu}, C.~Ederer, {N.A. Spaldin}, {R.R. Das}, {D.M. Kim}, {S.H. Baek},
  {C.B. Eom}, and R.~Ramesh, \emph{Nat. Materials}, vol.~5, p. 823, 2006.

\bibitem{wang2004b}
J.~Wang, H.~Zheng, Z.~Ma, S.~Prasertchoung, M.~Wuttig, R.~Droopad, J.~Yu,
  K.~Eisenbeiser, and R.~Ramesh, \emph{Appl. Phys. Lett.}, vol.~85, p. 2574,
  2004.

\bibitem{scott2000}
{J.F. Scott}, \emph{Ferroelectric memories}, Springer, Ed., 2000.

\bibitem{fujitsu2006}
Corporate press release, Dec 15, 2006, from Fujitsu America, Sunnydale,
  California.

\bibitem{ghosh2005}
S.~Ghosh, V.~Sih, {W.H. Lau}, {D.D. Awschalom}, {S.-Y. Bae}, S.~Wang,
  S.~Vaidya, and G.~Chapline, \emph{Appl. Phys. Lett.}, vol.~86, p. 232507,
  2005.

\bibitem{thiele2005}
C.~Thiele, K.~D\"orr, S.~F\"ahler, L.~Schultz, {D.C. Meyer}, {A.A. Levin}, and
  P.~Paufler, \emph{Appl. Phys. Lett.}, vol.~87, p. 262502, 2005.

\bibitem{sun99a}
{J.Z. Sun}, \emph{J. Magn. Magn. Mater.}, vol. 202, p. 157, 1999.

\bibitem{thiele2006}
C.~Thiele, K.~D\"orr, O.~Bilani, J.~R\"odel, and L.~Schultz, condmat/0600760.

\bibitem{eerenstein2006b}
W.~Eerenstein, M.~Wiora, {J.L. Prieto}, {J.F. Scott}, and {N.D. Mathur},
  condmat/0609209.

\bibitem{duan2006}
{C.G. Duan}, {S.S. Jaswal}, and {E.Y. Tsymbal}, \emph{Phys. Rev. Lett.},
  vol.~97, p. 047201, 2006.

\bibitem{ohtomo2002}
A.~Ohtomo, {D.A. Muller}, {J.L Grazul}, and {H.Y. Wang}, \emph{Nature}, vol.
  419, p. 378, 2006.

\bibitem{takizawa2006}
M.~Takizawa, H.~Wadati, K.~Tanaka, M.~Hashimoto, T.~Yoshida, A.~Fujimori,
  A.~Chikamatsu, H.~Kumigashira, M.~Oshima, K.~Shibuya, T.~Mihara, Y.~Ohnishi,
  M.~Lippmaa, M.~Kawasaki, H.~Koinuma, S.~Okamoto, and {A.J. Millis},
  \emph{Phys. Rev. Lett.}, vol.~97, p. 057601, 2006.

\bibitem{okamoto2004}
S.~Okamoto and {A.J. Millis}, \emph{Nature}, vol. 427, p. 630, 2004.

\bibitem{okamoto2005}
------, condmat/0506172.

\bibitem{okamoto2006}
S.~Okamoto, {A.J. Millis}, and {N.A. Spaldin}, condmat/0601081.

\bibitem{kancharla2006}
{S.S. Kancharla} and {E. Dagotto}, condmat/0606621.

\bibitem{thulasi2006}
S.~Thulasi and S.~Satpadhy, \emph{Phys. Rev. B}, vol.~73, p. 125307, 2006.

\bibitem{lee2006}
{W.-C. Lee} and {A.H. MacDonald}, condmat/0606200.

\bibitem{ohtomo2004}
A.~Ohtomo and {H.Y. Hwang}, \emph{Nature}, vol. 427, p. 423, 2004.

\bibitem{herranz2006c}
G.~Herranz, M.~Basletic, M.~Bibes, C.~Carretero, E.~Tafra, E.~Jacquet,
  K.~Bouzehouane, C.~Deranlot, {J.-L. Maurice}, A.~Hamzic, {J.-P. Contour},
  A.~Barth\'el\'emy, and A.~Fert, condmat/0606182.

\bibitem{tufte67}
{O.N. Tufte} and {P.W. Chapman}, \emph{Phys. Rev.}, vol. 155, p. 796, 1967.

\bibitem{herranz2006b}
G.~Herranz, M.~Basletic, M.~Bibes, R.~Ranchal, A.~hamzic, E.~Tafra,
  K.~Bouzehouane, E.~Jacquet, {J.-P. Contour}, A.~Barth\'el\'emy, and A.~Fert,
  \emph{Phys. Rev. B}, vol.~73, p. 064403, 2006.

\bibitem{tsukazaki2006}
A.~Tsukazaki, A.~Ohtomo, and M.~Kawasaki, \emph{Appl. Phys. Lett.}, vol.~88, p.
  152106, 2006.

\bibitem{ueno2003}
K.~Ueno, {I.H. Inoue}, H.~Akoh, M.~Kawasaki, Y.~Tokur, and H.~Takagi,
  \emph{Appl. Phys. Lett.}, vol.~83, p. 1755, 2003.

\bibitem{takahashi2006}
{K.S. Takahashi}, M.~Gabay, D.~Jaccard, K.~Shibuya, T.~Ohnishi, M.~Lippmaa, and
  {J.-M. Triscone}, \emph{Nature}, vol. 441, p. 195, 2006.

\bibitem{sui2004}
{Y.C. Sui}, R.~Skomski, {K.D. Sorge}, and {D.J. Sellmyer}, \emph{Appl. Phys.
  Lett.}, vol.~84, p. 1525, 2004.

\bibitem{liao2006}
{Z.-M. Liao}, {Y.-D. Li}, J.~Xu, {J.-M. Zhang}, K.~Xia, and D.~Yu,
  \emph{Nanolett.}, vol.~6, p. 1087, 2006.

\bibitem{curiale2005}
J.~Curiale, {R.D. Sanchez}, {H.E. Troiani}, {A.G. Leyva}, and P.~Levy,
  \emph{Appl. Phys. Lett.}, vol.~87, p. 043113, 2005.

\bibitem{wu2005}
D.~Wu, Y.~Chen, J.~Liu, X.~Zhao, A.~Li, and N.~Ming, \emph{Appl. Phys. Lett.},
  vol.~87, p. 112501, 2005.

\bibitem{krusin2004}
{L. Krusin-Elbaum}, {D.N. Newns}, H.~Zeng, V.~Derycke, {J.Z. Sun}, and
  R.~Sandstrom, \emph{Nature}, vol. 431, p. 672, 2004.

\bibitem{sahoo2005}
S.~Sahoo, T.~Kontos, J.~Furer, C.~Hoffmann, M.~Gr\"aber, A.~Cottey, and
  C.~Sch\"onberger, \emph{Nat. Phys.}, vol.~1, p.~99, 2005.

\bibitem{hueso2006}
{L.E. Hueso}, G.~Burnell, {J.L. Prieto}, L.~Granja, and C.~Bell, \emph{Appl.
  Phys. Lett.}, vol.~88, p. 083120, 2006.

\bibitem{hueso2006a}
{L.E. Hueso}, {J.M. Pruneda}, V.~Ferrari, G.~Burnell, {J.P. Vald\'es-Herrera},
  {B.D. Simmons}, {P.B. Littlewood}, E.~Artacho, and {N.D. Mathur},
  \emph{Nature}, vol. 445, p. 410, 2007.

\end{thebibliography}
\end{document}